\documentclass[aps,showpacs,preprintnumbers,nofootinbibt,preprint]{revtex4}
\usepackage{graphicx}
\usepackage{mathrsfs}

\begin{document}

\title{Thin accretion disks onto brane world black holes}
\author{C. S. J. Pun$^1$}
\email{jcspun@hkucc.hku.hk}
\author{Z. Kov\'{a}cs$^{2,3}$}
\email{zkovacs@mpifr-bonn.mpg.de}
\author{T. Harko$^1$}
\email{harko@hkucc.hku.hk}
\affiliation{$^{1}$Department of Physics and Center for Theoretical and Computational
Physics, The University of Hong Kong, Pok Fu Lam Road, Hong Kong}
\affiliation{$^{2}$Max-Planck-Institut f\"{u}r Radioastronomie, Auf dem H\"{u}gel 69,
53121 Bonn, Germany}
\affiliation{$^{3}$Department of Experimental Physics, University of Szeged, D\'{o}m T%
\'{e}r 9, Szeged 6720, Hungary}
\date{\today }

\begin{abstract}
The braneworld description of our universe entails a large extra dimension
and a fundamental scale of gravity that might be lower by several orders of
magnitude as compared to the Planck scale. An interesting consequence of the
braneworld scenario is in the nature of the vacuum
solutions of the brane gravitational field equations, with properties quite
distinct as compared to the standard black hole solutions of general
relativity. One possibility of observationally discriminating between
different types of black holes is the study of the emission properties of
the accretion disks. In the present paper we obtain the energy flux, the
emission spectrum and accretion efficiency from the accretion disks around
several classes of static and rotating brane world black holes, and
we compare them to the general relativistic case. Particular signatures can
appear in the electromagnetic spectrum, thus leading to the possibility of
directly testing extra-dimensional physical models by using astrophysical
observations of the emission spectra from accretion disks.
\end{abstract}

\maketitle




\section{Introduction}

The idea, proposed in \cite{RS99a,RS99b}, that our four-dimensional Universe
might be a three-brane, embedded in a five-dimensional space-time (the
bulk), has attracted a considerable interest in the past few years.
According to the brane-world scenario, the physical fields (electromagnetic,
Yang-Mills etc.) in our four-dimensional Universe are confined to the three
brane. These fields are assumed to arise as fluctuations of branes in string
theories. Only gravity can freely propagate in both the brane and bulk
space-times, with the gravitational self-couplings not significantly
modified. This model originated from the study of a single $3$-brane
embedded in five dimensions, with the $5D$ metric given by $%
ds^{2}=e^{-f(y)}\eta _{\mu \nu }dx^{\mu }dx^{\nu }+dy^{2}$, which, due to
the appearance of the warp factor, could produce a large hierarchy between
the scale of particle physics and gravity. Even if the fifth dimension is
uncompactified, standard $4D$ gravity is reproduced on the brane. Hence this
model allows the presence of large, or even infinite non-compact extra
dimensions. Our brane is identified to a domain wall in a $5$-dimensional
anti-de Sitter space-time. For a review of the brane world models see \cite{Mart04}.

The braneworld description of our universe entails a large extra dimension
and a fundamental scale of gravity that might be lower by several orders of
magnitude compared to the Planck scale \cite{RS99a,RS99b}. Due to the
correction terms coming from the extra dimensions, significant deviations
from the Einstein theory occur in brane world models at very high energies
\cite{SMS00,SSM00}. Gravity is largely modified at the electro-weak scale of
$1$ TeV. The cosmological and astrophysical implications of the brane world
theories have been extensively investigated in the physical literature~\cite%
{all2,sol}.

Several classes of spherically symmetric solutions of the static
gravitational field equations in the vacuum on the brane have been obtained
in~\cite{Ha03,Ma04,Ha04,Ha05}. As a possible physical application of these
solutions the behavior of the angular velocity $v_{tg}$ of the test
particles in stable circular orbits has been considered~\cite{Ma04,Ha04,Ha05}%
. The general form of the solution, together with two constants of
integration, uniquely determines the rotational velocity of the particle. In
the limit of large radial distances, and for a particular set of values of
the integration constants the angular velocity tends to a constant value.
This behavior is typical for massive particles (hydrogen clouds) outside
galaxies, and is usually explained by postulating the existence of the dark
matter. The exact galactic metric, the dark radiation, the dark pressure and
the lensing in the flat rotation curves region in the brane world scenario
has been obtained in \cite{Ha05}.

For standard general relativistic spherical compact objects the exterior
space-time is described by the Schwarzschild metric. In the five dimensional
brane world models, the high energy corrections to the energy density,
together with the Weyl stresses from bulk gravitons, imply that on the brane
the exterior metric of a static star is no longer the Schwarzschild metric
\cite{Da00}. The presence of the Weyl stresses also means that the matching
conditions do not have a unique solution on the brane; the knowledge of the
five-dimensional Weyl tensor is needed as a minimum condition for uniqueness.

It is known that the Einstein field equations in five dimensions admit more
general spherically symmetric black holes on the brane than four-dimensional
general relativity. Hence an interesting consequence of the braneworld
scenario is in the nature of the spherically symmetric vacuum solutions to
the brane gravitational field equations, which could represent black holes
with properties quite distinct as compared to ordinary black holes in four
dimensions. Such black holes are likely to have very diverse cosmological
and astrophysical signatures. Static, spherically symmetric exterior vacuum
solutions of the brane world models have been proposed first in \cite{Da00}
and in \cite{GeMa01}. The first of these solutions, obtained in \cite{Da00},
has the mathematical form of the Reissner-Nordstr\"{o}m solution of the
standard general relativity, in which a tidal Weyl parameter plays the role
of the electric charge of the general relativistic solution. The solution
has been obtained by imposing the null energy condition on the 3-brane for a
bulk having non zero Weyl curvature, and it can be matched to the interior
solution corresponding to a constant density brane world star. A second
exterior solution, which also matches a constant density interior, has been
derived in \cite{GeMa01}.

Two families of analytic solutions of the spherically symmetric vacuum brane
world model equations (with $g_{tt}\neq -1/g_{rr}$), parameterized by the
ADM mass and a PPN parameter $\beta $ have been obtained in \cite{cfm02}.
Non-singular black-hole solutions in the brane world model have been
considered in \cite{Da03}, by relaxing the condition of the zero scalar
curvature but retaining the null energy condition. The four-dimensional
Gauss and Codazzi equations for an arbitrary static spherically symmetric
star in a Randall--Sundrum type II brane world have been completely solved
on the brane in \cite{Vi03}. The on-brane boundary can be used to determine
the full $5$-dimensional space-time geometry. The procedure can be
generalized to solid objects such as planets.

A method to extend into the bulk asymptotically flat static spherically
symmetric brane-world metrics has been proposed in \cite{Ca03}. The exact
integration of the field equations along the fifth coordinate was done by
using the multipole ($1/r$) expansion. The results show that the shape of
the horizon of the brane black hole solutions is very likely a flat
``pancake'' for astrophysical sources.

The general solution to the trace of the 4-dimensional Einstein equations
for static, spherically symmetric configurations has been used as a basis
for finding a general class of black hole metrics, containing one arbitrary
function $g_{tt}=A(r)$, which vanishes at some $r=r_{h}>0$ (the horizon
radius) in \cite{BMD03}. Under certain reasonable restrictions, black hole
metrics are found, with or without matter. Depending on the boundary
conditions the metrics can be asymptotically flat, or have any other
prescribed asymptotic. The exact stationary and axisymmetric solutions
describing charged rotating black holes localized on a 3-brane in the
Randall-Sundrum braneworld were studied in \cite{AlGu05}. \ By taking the
metric on the brane to be of the Kerr-Schild form it can be shown that the
Kerr-Newman solution of ordinary general relativity in which the electric
charge is superseded by a tidal charge satisfies a closed system of the
effective gravitational field equations on the brane. The negative tidal
charge may provide a mechanism for spinning up the black hole so that its
rotation parameter exceeds its mass. For a review of the black hole
properties and of the lensing in the brane world models see \cite{MaMu05}.

It is generally expected that most of the astrophysical objects grow
substantially in mass via accretion. Recent observations suggest that around
most of the active galactic nuclei (AGN's) or black hole candidates there
exist gas clouds surrounding the central compact object, and an associated
accretion disc, on a variety of scales from a tenth of a parsec to a few
hundred parsecs \cite{UrPa95}. These clouds are assumed to form a
geometrically and optically thick torus (or warped disc), which absorbs most
of the ultraviolet radiation and the soft X-rays. The gas exists in either
the molecular or the atomic phase. The most powerful evidence for the
existence of super massive black holes comes from the very-long baseline interferometry (VLBI) imaging of
molecular $\mathrm{H_{2}O}$ masers in the active galaxy NGC 4258 \cite%
{Miyo95}. This imaging, produced by Doppler shift measurements assuming
Keplerian motion of the masering source, has allowed a quite accurate
estimation of the central mass, which has been found to be a $3.6\times
10^{7}M_{\odot }$ super massive dark object, within $0.13$ parsecs. Hence,
important astrophysical information can be obtained from the observation of
the motion of the gas streams in the gravitational field of compact objects.

The determination of the accretion rate for an astrophysical object can give
a strong evidence for the existence of a surface of the object. A model in
which Sgr A*, the $3.7\times 10^6 M_{\odot }$ super massive black hole
candidate at the Galactic center, may be a compact object with a thermally
emitting surface was considered in \cite{BrNa06}. For very compact surfaces
within the photon orbit, the thermal assumption is likely to be a good
approximation because of the large number of rays that are strongly
gravitationally lensed back onto the surface. Given the very low quiescent
luminosity of Sgr A* in the near-infrared, the existence of a hard surface,
even in the limit in which the radius approaches the horizon, places a
severe constraint on the steady mass accretion rate onto the source, ${\dot M%
}\le 10^{-12} M_{\odot}$ yr$^{-1}$. This limit is well below the minimum
accretion rate needed to power the observed submillimeter luminosity of Sgr
A*, ${\dot M}\ge 10^{-10} M_{\odot}$ yr$^{}$.

Thus, from the determination of the accretion rate it follows that Sgr A*
does not have a surface, that is, it must have an event horizon. Therefore
the study of the accretion processes by compact objects is a powerful
indicator of their physical nature.

The first comprehensive theory of accretion disks around black holes was
constructed in \cite{ShSu73}. This theory was extended to the general
relativistic models of the mass accretion onto rotating black holes in \cite%
{NoTh73}. These pioneering works developed thin steady-state accretion
disks, where the accreting matter moves in Keplerian orbits. The
hydrodynamical equilibrium in the disk is maintained by an efficient cooling
mechanism via radiation transport. The photon flux emitted by the disk
surface was studied under the assumption that the disk emits a black body
radiation. The properties of radiant energy flux over the thin accretion
disks were further analyzed in \cite{PaTh74} and in \cite{Th74}, where the
effects of the photon capture by the hole on the spin evolution were
presented as well. In these works the efficiency with which black holes
convert rest mass into outgoing radiation in the accretion process was also
computed.

The emissivity properties of the accretion disks have also been investigated
for exotic central objects recently, such as quark, boson or fermion stars
for both rotating and non-rotating cases \cite{Bom,To02,YuNaRe04}, as well as for the modified $f(R)$ type theories of gravity \cite{mod}. The radiation power per unit area, the temperature of the disk and the spectrum
of the emitted radiation were given, and compared with the case of a
Schwarzschild black hole of an equal mass.

It is the purpose of the present paper to study the matter accretion by
brane world black holes. By using the general formalism of accretion we
analyze the accretion process for several black hole type solutions of the
gravitational field equations on the brane, both non-rotating and rotating,
which have been previously obtained. Particular signatures can appear in the
electromagnetic spectrum, thus leading to the possibility of directly
testing extra-dimensional physical models by using astrophysical
observations of the emission spectra from accretion disks.

The present paper is organized as follows. We review the field equations of
the brane world models and the static, spherically symmetric solutions of the field
equations as well as the rotating ones in Section II. The thin accretion disks onto black holes are
briefly described in Section III. In Section IV we consider the radiation
flux, spectrum and efficiency of thin accretion disks onto several classes
of brane world black holes. We discuss and conclude our results in Section V.

\section{The gravitational field equations in the brane world models}

\label{field}

In the present Section we briefly describe the basic mathematical formalism
of the brane world models, and we present the spherically symmetric static
vacuum field equations. The solutions of the vacuum field equations on the
brane physically describe the brane world black holes.

\subsection{The gravitational field equations on the brane}

We start by considering a five dimensional (5D) spacetime (the bulk), with a
single four-dimensional (4D) brane, on which matter is confined. The 4D
brane world $({}^{(4)}M,g_{\mu \nu })$ is located at a hypersurface $\left(
B\left( X^{A}\right) =0\right) $ in the 5D bulk spacetime $%
({}^{(5)}M,g_{AB}) $, of which coordinates are described by $%
X^{A},A=0,1,...,4$. The induced 4D coordinates on the brane are $x^{\mu
},\mu =0,1,2,3$. In the present paper the capital Latin indices $A$, $%
B,...,I,J,...$ take values in the range $0,1,...,4$, while the Greek indices
run in the range $0,...,3$.

The action of the system is given by $S=S_{bulk}+S_{brane}$, where\bigskip
\[
S_{bulk}=\int_{{}^{(5)}M}\sqrt{-{}^{(5)}g}\left[ \frac{1}{2k_{5}^{2}}{}%
^{(5)}R+{}^{(5)}L_{m}+\Lambda _{5}\right] d^{5}X,
\]%
and
\begin{equation}
S_{brane}=\int_{{}^{(4)}M}\sqrt{-{}^{(5)}g}\left[ \frac{1}{k_{5}^{2}}K^{\pm
}+L_{brane}\left( g_{\alpha \beta },\psi \right) +\lambda _{b}\right] d^{4}x,
\end{equation}%
where $k_{5}^{2}=8\pi G_{5}$ is the 5D gravitational constant, ${}^{(5)}R$
and ${}^{(5)}L_{m}$ are the 5D scalar curvature and the matter Lagrangian in
the bulk, $L_{brane}\left( g_{\alpha \beta },\psi \right) $ is the 4D
Lagrangian, which is given by a generic functional of the brane metric $%
g_{\alpha \beta }$ and of the matter fields $\psi $, $K^{\pm }$ is the trace
of the extrinsic curvature on either side of the brane, and $\Lambda _{5}$
and $\lambda _{b}$ (the constant brane tension) are the negative vacuum
energy densities in the bulk and on the brane, respectively~\cite{SMS00}.

The Einstein field equations in the bulk are given by~\cite{SMS00}
\begin{equation}
{}^{(5)}G_{IJ}=k_{5}^{2} {}^{(5)}T_{IJ},\qquad {}^{(5)}T_{IJ}=-\Lambda _{5}
{}^{(5)}g_{IJ}+\delta(B)\left[-\lambda_{b} {}^{(5)}g_{IJ}+T_{IJ}\right] ,
\end{equation}
where ${}^{(5)}T_{IJ}\equiv - 2\delta {}^{(5)}L_{m}/\delta {}^{(5)}g^{IJ}
+{}^{(5)}g_{IJ} {}^{(5)}L_{m}$, is the energy-momentum tensor of bulk matter
fields, while $T_{\mu \nu }$ is the energy-momentum tensor localized on the
brane and which is defined by $T_{\mu \nu }\equiv -2\delta L_{brane}/\delta
g^{\mu \nu }+g_{\mu \nu }\text{ }L_{brane}$.

The delta function $\delta \left( B\right) $ denotes the localization of
brane contribution. In the 5D spacetime a brane is a fixed point of the $%
Z_{2}$ symmetry. The basic equations on the brane are obtained by
projections onto the brane world. The induced 4D metric is $%
g_{IJ}={}^{(5)}g_{IJ}-n_{I}n_{J}$, where $n_{I}$ is the space-like unit
vector field normal to the brane hypersurface ${}^{(4)}M$. The unit vector
field satisfies the condition $n_{I}n^{I}=-1$. In the following we assume $%
{}^{(5)}L_{m}=0$. In the brane world models only gravity can probe the extra
dimensions.

Assuming a metric of the form $ds^{2}=(n_{I}n_{J}+g_{IJ})dx^{I}dx^{J}$, with
$n_{I}dx^{I}=d\chi $ the unit normal to the $\chi =\mathrm{constant}$
hypersurfaces and $g_{IJ}$ the induced metric on $\chi =\mathrm{constant}$
hypersurfaces, the effective 4D gravitational equation on the brane takes
the form~\cite{SMS00}:
\begin{equation}
G_{\mu \nu }=-\Lambda g_{\mu \nu }+k_{4}^{2}T_{\mu \nu }+k_{5}^{4}S_{\mu \nu
}-E_{\mu \nu },  \label{Ein}
\end{equation}
where $S_{\mu \nu }$ is the local quadratic energy-momentum correction
\begin{equation}
S_{\mu \nu }=\frac{1}{12}TT_{\mu \nu }-\frac{1}{4}T_{\mu }{}^{\alpha }T_{\nu
\alpha }+\frac{1}{24}g_{\mu \nu }\left( 3T^{\alpha \beta }T_{\alpha \beta
}-T^{2}\right) ,
\end{equation}
and $E_{\mu \nu }$ is the non-local effect from the free bulk gravitational
field, the transmitted projection of the bulk Weyl tensor $C_{IAJB}$, $%
E_{IJ}=C_{IAJB}n^{A}n^{B}$, with the property $E_{IJ}\rightarrow E_{\mu \nu
}\delta _{I}^{\mu }\delta _{J}^{\nu }\quad $as$\quad \chi \rightarrow 0$. We
have also denoted $k_{4}^{2}=8\pi G$, with $G$ the usual 4D gravitational
constant.

The 4D cosmological constant, $\Lambda $, and the 4D coupling constant, $%
k_{4}$, are related by $\Lambda =k_{5}^{2}(\Lambda _{5}+k_{5}^{2}\lambda
_{b}^{2}/6)/2$ and $k_{4}^{2}=k_{5}^{4}\lambda _{b}/6$, respectively. In the
limit $\lambda _{b}^{-1}\rightarrow 0$ we recover standard general
relativity \cite{SMS00}.

The Einstein equation in the bulk and the Codazzi equation also imply the
conservation of the energy-momentum tensor of the matter on the brane, $%
D_{\nu }T_{\mu }{}^{\nu }=0$, where $D_{\nu }$ denotes the brane covariant
derivative. Moreover, from the contracted Bianchi identities on the brane it
follows that the projected Weyl tensor obeys the constraint $D_{\nu }E_{\mu
}{}^{\nu }=k_{5}^{4}D_{\nu }S_{\mu }{}^{\nu }$.

The symmetry properties of $E_{\mu \nu }$ imply that in general we can
decompose it irreducibly with respect to a chosen $4$-velocity field $%
u^{\mu} $ as $E_{\mu \nu }=-k^{4}\left[ U\left( u_{\mu }u_{\nu} +\frac{1}{3}%
h_{\mu \nu }\right) +P_{\mu \nu }+2Q_{(\mu }u_{\nu)}\right]$, where $%
k=k_{5}/k_{4}$, $h_{\mu \nu }=g_{\mu \nu }+u_{\mu }u_{\nu }$ projects
orthogonal to $u^{\mu }$, the ``dark radiation'' term $U=-k^{-4}E_{\mu \nu
}u^{\mu }u^{\nu }$ is a scalar, $Q_{\mu }=k^{-4}h_{\mu }^{\alpha }E_{\alpha
\beta }u^{\beta }$ is a spatial vector and $P_{\mu \nu }=-k^{-4}\left[
h_{(\mu }\text{ }^{\alpha }h_{\nu )}\text{ }^{\beta }-\frac{1}{3}h_{\mu \nu
}h^{\alpha \beta }\right] E_{\alpha \beta }$ is a spatial, symmetric and
trace-free tensor~\cite{Mart04}.

In the case of the vacuum state we have $\rho =p=0$, $T_{\mu \nu }\equiv 0$,
and consequently $S_{\mu \nu }\equiv 0$. Therefore the field equation
describing a static brane takes the form
\begin{equation}
R_{\mu \nu }=-E_{\mu \nu }+\Lambda g_{\mu \nu },
\end{equation}
with the trace $R$ of the Ricci tensor $R_{\mu \nu }$ satisfying the
condition $R=R_{\mu }^{\mu }=4\Lambda $.

In the vacuum case $E_{\mu \nu }$ satisfies the constraint $D_{\nu }E_{\mu
}{}^{\nu }=0$. In an inertial frame at any point on the brane we have $%
u^{\mu }=\delta _{0}^{\mu }$ and $h_{\mu \nu }=\mathrm{diag}(0,1,1,1)$. In a
static vacuum $Q_{\mu }=0$ and the constraint for $E_{\mu \nu }$ takes the
form~ \cite{GeMa01}
\begin{equation}
\frac{1}{3}D_{\mu }U+\frac{4}{3}UA_{\mu }+D^{\nu }P_{\mu \nu }+A^{\nu
}P_{\mu \nu }=0,
\end{equation}
where $A_{\mu }=u^{\nu }D_{\nu }u_{\mu }$ is the 4-acceleration. In the
static spherically symmetric case we may chose $A_{\mu }=A(r)r_{\mu }$ and $%
P_{\mu \nu }=P(r)\left( r_{\mu }r_{\nu }-\frac{1}{3}h_{\mu \nu }\right) $,
where $A(r)$ and $P(r)$ (the ``dark pressure'') are some scalar functions of
the radial distance~$r$, and~$r_{\mu }$ is a unit radial vector~\cite{Da00}.

\subsection{Static and spherically symmetric brane world black holes}

Static black holes are described by the static and spherically symmetric
metric given by
\begin{equation}
ds^{2}=-e^{\nu (r)}dt^{2}+e^{\lambda (r)}dr^{2}+r^{2}\left( d\theta
^{2}+\sin ^{2}\theta d\phi ^{2}\right) .  \label{metr1}
\end{equation}

With the metric given by~(\ref{metr1}) the gravitational field equations and
the effective energy-momentum tensor conservation equation in the vacuum
take the form~\cite{Ha03,Ma04}
\begin{equation}
-e^{-\lambda }\left( \frac{1}{r^{2}}-\frac{\lambda ^{\prime }}{r}\right) +%
\frac{1}{r^{2}}=3\alpha U+\Lambda ,  \label{f1}
\end{equation}
\begin{equation}
e^{-\lambda }\left( \frac{\nu ^{\prime }}{r}+\frac{1}{r^{2}}\right) -\frac{1%
}{r^{2}}=\alpha \left( U+2P\right) -\Lambda ,  \label{f2}
\end{equation}
\begin{equation}
\frac{1}{2}e^{-\lambda }\left( \nu ^{\prime \prime }+\frac{\nu ^{\prime 2}}{2%
}+\frac{\nu ^{\prime }-\lambda ^{\prime }}{r}-\frac{\nu ^{\prime }\lambda
^{\prime }}{2}\right) =\alpha \left( U-P\right) -\Lambda ,  \label{f3}
\end{equation}
\begin{equation}
\nu ^{\prime }=-\frac{U^{\prime }+2P^{\prime }}{2U+P}-\frac{6P}{r\left(
2U+P\right) },  \label{f4}
\end{equation}
where $^{\prime }=d/dr$, and we have denoted $\alpha =16\pi G/k^{4}\lambda
_{b}$.

The field equations~(\ref{f1})--(\ref{f3}) can be interpreted as describing
an anisotropic ''matter distribution'', with the effective energy density $%
\rho ^{\mathrm{eff}}$, radial pressure $P^{\mathrm{eff}}$ and orthogonal
pressure $P_{\perp }^{\mathrm{eff}}$, respectively, so that $\rho ^{\mathrm{%
eff}}=3\alpha U+\Lambda $, $P^{\mathrm{eff}}=\alpha U+2\alpha P-\Lambda $
and $P_{\perp }^{\mathrm{eff}}=\alpha U-\alpha P-\Lambda $, respectively,
which gives the condition $\rho ^{\mathrm{eff}}-P^{\mathrm{eff}}-2P_{\perp
}^{\mathrm{eff}}=4\Lambda =\mathrm{constant}$. This is expected for the
`radiation' like source, for which the projection of the bulk Weyl tensor is
trace-less, $E_{\mu }^{\mu }=0$.

Eq.~(\ref{f1}) can immediately be integrated to give
\begin{equation}
e^{-\lambda }=1-\frac{C_{1}}{r}-\frac{GM_{U}\left( r\right) }{r}-\frac{%
\Lambda }{3}r^{2},  \label{m1}
\end{equation}
where $C_{1}$ is an arbitrary constant of integration, and we denoted $%
GM_{U}\left( r\right) =3\alpha \int_{0}^{r}U(r)r^{2}dr$.

The function $M_U$ is the gravitational mass corresponding to the dark
radiation term (the dark mass). For $U=0$ the metric coefficient given by
Eq.~(\ref{m1}) must tend to the standard general relativistic Schwarzschild
metric coefficient, which gives $C_{1}=2GM$, where $M = \mathrm{constant}$
is the baryonic (usual) mass of the gravitating system.

By substituting $\nu ^{\prime }$ given by Eq.~(\ref{f4}) into Eq.~(\ref{f2}%
), and with the use of Eq.~(\ref{m1}), we obtain the following system of
differential equations satisfied by the dark radiation term $U$, the dark
pressure $P$ and the dark mass $M_{U}$, describing the vacuum gravitational
field, exterior to a massive body, in the brane world model \cite{Ha03}:
\begin{equation}
\frac{dM_{U}}{dr}=\frac{3\alpha }{G}r^{2}U.  \label{e2}
\end{equation}%
\begin{equation}
\frac{dU}{dr}=-\frac{\left( 2U+P\right) \left[ 2GM+GM_{U}-\frac{2}{3}\Lambda
r^{3}+\alpha \left( U+2P\right) r^{3}\right] }{r^{2}\left( 1-\frac{2GM}{r}-%
\frac{M_{U}}{r}-\frac{\Lambda }{3}r^{2}\right) }-2\frac{dP}{dr}-\frac{6P}{r},
\label{e1}
\end{equation}

To close the system a supplementary functional relation between one of the
unknowns $U$, $P$ and $M_{U}$ is needed. Generally, this equation of state
is given in the form $P=P(U)$. Once this relation is known, Eqs.~(\ref{e2}%
)--(\ref{e1}) give a full description of the geometrical properties of the
vacuum on the brane.

In the following we will restrict our analysis to the case $\Lambda =0$.
Then the system of equations~(\ref{e2}) and~(\ref{e1}) can be transformed to
an autonomous system of differential equations by means of the
transformations $q=2GM/r+GM_{U}/r$, $\mu =3\alpha r^{2}U$, $p=3\alpha r^{2}P$%
, $\theta =\ln r$ where $\mu $ and $p$ are the ``reduced'' dark radiation
and pressure, respectively. With the use of the new variables, Eqs.~(\ref{e2}%
) and~(\ref{e1}) become
\begin{equation}
\frac{dq}{d\theta }=\mu -q,  \label{aut1}
\end{equation}
\begin{equation}
\frac{d\mu }{d\theta }=-\frac{\left( 2\mu +p\right) \left[ q+\frac{1}{3}%
\left( \mu +2p\right) \right] }{1-q}-2\frac{dp}{d\theta }+2\mu -2p.
\label{aut2}
\end{equation}

Eqs.~(\ref{e2}) and~(\ref{e1}), or, equivalently,~(\ref{aut1}) and~(\ref%
{aut2}), are called the structure equations of the vacuum on the brane \cite%
{Ha03}. In order to close the system of equations (\ref{aut1}) and~(\ref%
{aut2}) an \textquotedblleft equation of state\textquotedblright\ $p=p\left(
\mu \right) $, relating the reduced dark radiation and the dark pressure
terms, is needed. Once the equation of state is known, exact vacuum
solutions of the gravitational field equations on the brane can be obtained.
The opposite procedure can also be followed, that is, by specifying the
functional form of the metric tensor, the dark radiation and the dark
pressure can be obtained from the field equations. Therefore, several exact
solutions of the gravitational field equations on the brane can be obtained
\cite{Da00,cfm02,BMD03}.

\subsection{Rotating brane world black holes}

In order to study rotating black hole solutions on the brane it is
convenient to assume that the axisymmetric and stationary metric is of the
Kerr-Schild form, and can be expressed in the form of its linear
approximation around the flat metric, $ds^{2}=\left( ds^{2}\right)
_{flat}+H\left( l_{\mu }dx^{\mu }\right) ^{2}$, where $l_{\mu }$ is a null,
geodesic vector field in both the flat and full metrics, and $H$ is an
arbitrary scalar function. By introducing a set of coordinates $y^{\mu
}=\left\{ u,r,\theta ,\phi \right\} $, the metric can be written in the
alternative form \cite{AlGu05}

\begin{equation}\label{rotmetr}
ds^{2}=\left[ -\left( du+dr\right) ^{2}+dr^{2}+\Sigma d\theta ^{2}+\left(
r^{2}+a^{2}\right) \sin ^{2}\theta d\phi ^{2}+2a\sin ^{2}\theta drd\phi
+H\left( du-a\sin ^{2}\theta d\phi \right) ^{2}\right] ,
\end{equation}%
where $H=H\left( r,\theta \right) $, the parameter $a$ is related to the
angular momentum of the black hole, and the quantity $\Sigma $ is defined as
$\Sigma =r^{2}+a^{2}\cos ^{2}\theta $. The condition $R=0$ gives for $H$ the
differential equation

\begin{equation}
\left( \frac{\partial ^{2}}{\partial r^{2}}+\frac{4r}{\Sigma }\frac{\partial
}{\partial r}+\frac{2}{\Sigma }\right) H=0,
\end{equation}%
with the general solution $H=\left( 2Mr-\beta \right) /\Sigma $, where $M$
and $\beta $ are arbitrary integration constants. By applying the
Boyer-Lindquist transformation $du=dt-\left( r^{2}+a^{2}\right) dr/\Delta $,
$d\phi =d\varphi -adr/\Delta $, where $\Delta =r^{2}+a^{2}-2Mr+\beta $, the
induced metric for a rotating black hole on the brane takes the form of the
Kerr-Newman solution of standard general relativity, describing a stationary
and axisymmetric charged black hole. The parameter $M$ can be interpreted as
the mass of the black hole, but since there is no electric charge on the
brane, the parameter $\beta $, the tidal charge parameter, which can take
both positive and negative values, carries the imprints of a non-local,
Coulomb type interaction from the bulk. The components of the projections of
the Weyl tensor from the bulk are given by $E_{t}^{t}=-E_{\varphi }^{\varphi
}=-\beta \left[ \Sigma -2\left( r^{2}+a^{2}\right) \right] /\Sigma ^{3}$, $%
E_{r}^{r}=-E_{\theta }^{\theta }=\beta /\Sigma ^{2}$, and $E_{\varphi
}^{t}=-2\beta a\left( r^{2}+a^{2}\right) \sin ^{2}\theta /\Sigma ^{3}$,
which shows a clear analogy with the energy-momentum tensor of a charged
rotating black hole in standard general relativity \cite{AlGu05}.

\section{Thin accretion disks onto black holes}

The theory of mass accretion around rotating black holes was developed for
the general relativistic case by Novikov and Thorne \cite{NoTh73}. They
extended the steady-state thin disk models introduced by Shakura and Sunyaev
\cite{ShSu73} to the curved space-time, by adopting the equatorial
approximation for the stationary and axisymmetric geometry. The time- and
space-like Killing vector fields $(\partial / \partial t)^{\mu}$ and $%
(\partial / \partial \phi)^{\mu}$ describe the symmetry properties of this
type of space-time, where $t$ and $r$ are the Boyer-Lyndquist time and
radial coordinates, respectively.

The horizontal size of the thin disk is negligible as compared to its vertical
extension, i.e, the disk height $H$, defined by the maximum half thickness of
the disk, is much smaller than any characteristic radii $r$ of the disk, $%
H<<r $. In the steady-state accretion disk models, the mass accretion rate $\dot{M%
}_{0} $ is supposed to be constant in time, and the physical quantities of
the accreting matter are averaged over a characteristic time scale, e.g. $%
\Delta t$, and over the azimuthal angle $\Delta \phi =2\pi $, for a total period of
the orbits and for the height $H$. The plasma moves in Keplerian orbits around
the black hole, with a rotational velocity $\Omega $, and the plasma
particles have a specific energy $\widetilde{E}$, and specific angular
momentum $\widetilde{L}$, which depend only on the radii of the orbits. The
particles are orbiting with the four-velocity $u^{\mu }$ in a disk having an
averaged surface density $\Sigma $. The accreting matter is modeled by an
anisotropic fluid source, where the density $\rho _{0}$ (the specific heat
is neglected), the energy flow vector $q^{\mu }$ and the stress tensor $%
t^{\mu \nu }$ are measured in the averaged rest-frame. The energy-momentum
tensor describing this source takes the form
\[
T^{\mu \nu }=\rho _{0}u^{\mu }u^{\nu }+2u^{(\mu }q^{\nu )}+t^{\mu \nu }\;,
\]%
where $u_{\mu }q^{\mu }=0$, $u_{\mu }t^{\mu \nu }=0$. The four-vectors of
the energy and of the angular momentum flux are defined by
\begin{equation}
-E^{\mu }\equiv T_{{}}^{\mu }{}_{\nu }(\partial /\partial t)^{\nu }\;\qquad
J^{\mu }\equiv T_{{}}^{\mu }{}_{\nu }(\partial /\partial \phi )^{\nu }\;,
\end{equation}%
respectively. The four dimensional conservation laws
of the rest mass, of the energy and of the angular momentum of the plasma
provide the structure equations of the thin disk. By integrating the
equation of the rest mass conservation, $\nabla _{\mu }(\rho _{0}u^{\mu })=0$, %
 it follows that the time averaged
 accretion rate  $\dot{M_{0}}$ is independent of the disk radius:
\begin{equation}
\dot{M_{0}}\equiv -2\pi r\Sigma u^{r}=\mbox{const}\;,  \label{conslawofM}
\end{equation}%
where a dot represents the derivative with respect to the time coordinate \cite{PaTh74}.
The averaged rest mass density  is defined by
\begin{equation}
\Sigma (r)=\int_{-H}^{H}\langle \rho _{0}\rangle dz,
\end{equation}%
where $\langle \rho _{0}\rangle $ is the rest mass density averaged  over $\Delta
t$ and $2\pi $. The conservation law $\nabla _{\mu }E^{\mu }=0$ of the
energy can be written in an integral form as
\begin{equation}
\lbrack \dot{M}_{0}\widetilde{E}-2\pi r\Omega W_{\phi }{}^{r}]_{,r}=4\pi
\sqrt{-g}F\widetilde{E}\;\;,  \label{conslawofE}
\end{equation}%
where a comma denotes the derivative with respect to the radial coordinate $r$. Eq.~(\ref{conslawofE}) shows the balance between the energy transported by the rest mass flow, the dynamical stresses in the disk, and the energy radiated away
from the surface of the disk, respectively. The torque $W_{\phi }{}^{r}$ in Eq.~(\ref{conslawofE}) is given by
\begin{equation}
W_{\phi }{}^{r}=\int_{-H}^{H}\langle t_{\phi }{}^{r}\rangle dz,
\end{equation}%
where $\langle t_{\phi }{}^{r}\rangle $ is the $\phi -r$ component of the stress
tensor, averaged over $\Delta t$ and over a $2\pi $ angle. The law of the angular momentum
conservation, $\nabla _{\mu }J^{\mu }=0$,  states in its integral form the
balance of the three forms of the angular momentum transport,
\begin{equation}
\lbrack \dot{M}_{0}\widetilde{L}-2\pi rW_{\phi }{}^{r}]_{,r}=4\pi \sqrt{-g}F%
\widetilde{L}\;\;.  \label{conslawofL}
\end{equation}

By eliminating $W_{\phi}{}^{r}$ from Eqs. (\ref{conslawofE}) and (\ref%
{conslawofL}), and by applying the universal energy-angular momentum relation
$dE=\Omega dJ$ for circular geodesic orbits in the form $\widetilde{E}%
_{,r}=\Omega\widetilde{L}_{,r}$, the flux of the
radiant energy over the disk can be expressed in terms of the specific energy, angular
momentum and the angular velocity of the black hole. Then the flux integral
leads to the expression of the energy flux $F(r)$, which is given by
\begin{equation}
F(r)=-\frac{\dot{M}_0}{4\pi\sqrt{-g}} \frac{\Omega_{,r}}{(\widetilde{E}%
-\Omega\widetilde{L})^{2}} \int_{r_{ms}}^{r}(\widetilde{E}-\Omega\widetilde{L%
})\widetilde{L}_{,r}dr\;,  \label{F}
\end{equation}
where the no-torque inner boundary conditions were also prescribed \cite%
{PaTh74}. This means that the torque vanishes at the inner edge of the disk,
since the matter at the marginally stable orbit $r_{ms}$ falls
freely into the black hole, and cannot exert considerable torque on the
disk. The latter assumption is valid as long as strong magnetic fields do
not exist in the plunging region, where matter falls into the hole.

Once the geometry of the space-time is known, we can derive the time
averaged radial distribution of photon emission for accretion disks around
black holes, and determine the efficiency of conversion of the rest mass into
outgoing radiation. After obtaining the radial dependence of
the angular velocity $\Omega $, of the specific energy $\widetilde{E}$ and
of the specific angular momentum $\widetilde{L}$ of the particles moving on
circular orbits around the black holes, respectively, we can compute the
flux integral (\ref{F}).

Let us consider an arbitrary stationary and axially symmetric geometry,
\begin{equation}
ds^{2}=g_{tt}dt^{2}+g_{t\phi }dtd\phi +g_{rr}dr^{2}+g_{\theta \theta
}d\theta ^{2}+g_{\phi \phi }d\phi ^{2}\;,  \label{ds2rcoappr}
\end{equation}%
where in the equatorial approximation ($|\theta -\pi /2|\ll 1$)the metric functions $g_{tt}$, $g_{t\phi }$, $g_{rr}$, $g_{\theta
\theta }$ and $g_{\phi \phi }$ depend only on the radial coordinate $r$. The geodesic
equations take the form
\begin{equation}
\left( \frac{dt}{d\tau }\right) ^{2}=\frac{\widetilde{E}g_{\phi \phi }+%
\widetilde{L}g_{t\phi }}{g_{t\phi }^{2}-g_{tt}g_{\phi \phi }},
\end{equation}
\begin{equation}
\left( \frac{d\phi }{d\tau }\right) ^{2}=-\frac{\widetilde{E}g_{t\phi }+%
\widetilde{L}g_{tt}}{g_{t\phi }^{2}-g_{tt}g_{\phi \phi }},
\end{equation}%
and
\begin{equation}
g_{rr}\left( \frac{dr}{d\tau }\right) ^{2}=V(r),
\end{equation}%
respectively, where $\tau $ is the affine parameter, and the potential term $V(r)$ is defined by
\begin{equation}
V(r)\equiv -1+\frac{\widetilde{E}^{2}g_{\phi \phi }+2\widetilde{E}%
\widetilde{L}g_{t\phi }+\widetilde{L}^{2}g_{tt\texttt{}}}{g_{t\phi
}^{2}-g_{tt}g_{\phi \phi }}\;.
\end{equation}

For circular orbits in the equatorial plane the following conditions must
hold%
\begin{equation}
V(r)=0,\qquad V_{,r}(r)=0.
\end{equation}

These conditions give the specific energy $\widetilde{E}$, the specific angular
momentum $\widetilde{L}$ and the angular velocity $\Omega $ of particles
moving on circular orbits around spinning general relativistic stars as%
\begin{eqnarray}
\widetilde{E} &=&-\frac{g_{tt}+g_{t\phi }\Omega }{\sqrt{-g_{tt}-2g_{t\phi
}\Omega -g_{\phi \phi }\Omega ^{2}}}\;,  \label{tildeE} \\
\widetilde{L} &=&\frac{g_{t\phi }+g_{\phi \phi }\Omega }{\sqrt{%
-g_{tt}-2g_{t\phi }\Omega -g_{\phi \phi }\Omega ^{2}}},  \label{tildeL} \\
\Omega  &=&\frac{d\phi }{dt}=\frac{-g_{t\phi ,r}+\sqrt{(g_{t\phi
,r})^{2}-g_{tt,r}g_{\phi \phi ,r}}}{g_{\phi \phi ,r}}\;.
\end{eqnarray}%
The marginally stable orbit around the central object are determined by
the condition%
\begin{equation}
V_{,rr}(r)=0.
\end{equation}

Let us represent the effective potential in the form $V(r)\equiv -1+f/g$,
where $f\equiv \widetilde{E}^{2}g_{\phi \phi }+2\widetilde{E}\widetilde{L}%
g_{t\phi }+\widetilde{L}^{2}g_{\phi \phi }$ and $g\equiv g_{t\phi
}^{2}-g_{tt}g_{\phi \phi }$, respectively. Then from the condition $%
V(r)=0$ we obtain first $-1+f/g=0$, which implies $f=g$.   From $%
V_{,r}(r)=0$ we obtain $\left( f_{,r}g-fg_{,r}\right) /g^{2}=0$, while $%
V_{,rr}(r)=0$ gives $g^{-1}(f_{,rr}-g_{,rr})=0$, since $V(r)=0$ and
$V_{,r}(r)=0$. If $g\neq 0$ we have%
\begin{equation}
\widetilde{E}^{2}g_{\phi \phi ,rr}+2\widetilde{E}\widetilde{L}g_{t\phi ,rr}+%
\widetilde{L}^{2}g_{tt ,rr}-(g_{t\phi }^{2}-g_{tt}g_{\phi \phi
})_{,rr}=0\;.  \label{stable}
\end{equation}

By inserting Eqs.~(\ref{tildeE})-(\ref{tildeL}) into Eq.~(\ref{stable}), and
solving the resulting equation for $r$, we obtain the marginally stable
orbits, once the metric coefficients $g_{tt}$, $g_{t\phi }$ and $%
g_{\phi \phi }$ are explicitly given.

In the case of a static and spherically symmetric geometry, given by Eq. (%
\ref{metr1}), the geodesic equations for particles orbiting in the
equatorial plane  take the form
\begin{equation}
e^{2\nu }\left( \frac{dt}{d\tau }\right) ^{2}=\widetilde{E}^{2},e^{(\nu
+\lambda )}\left( \frac{dr}{d\tau }\right) ^{2}+V_{eff}(r)=\widetilde{E}%
^{2},r^{4}\left( \frac{d\phi }{d\tau }\right) ^{2}=\widetilde{L}^{2},
\end{equation}%
and the effective potential can be written as
\begin{equation}
V(r)\equiv e^{\nu }\left( 1+\frac{\widetilde{L}}{r^{2}}^{2}\right) \;.
\label{V2}
\end{equation}%
The conditions for a stable particle orbit are again $V(r)=0$ and $%
V_{,r}(r)=0$. From these conditions we obtain
\begin{eqnarray}
\Omega  &=&\sqrt{\frac{\nu _{,r}e^{\nu }}{2r}}\;,  \label{Omega} \\
\widetilde{E} &=&\frac{e^{\nu }}{\sqrt{e^{\nu }-r^{2}\Omega ^{2}}}\;,
\label{E} \\
\widetilde{L} &=&\frac{r^{2}\Omega }{\sqrt{e^{\nu }-r^{2}\Omega ^{2}}}\;.
\label{L}
\end{eqnarray}%
At the marginally stable orbit (or the innermost stable circular orbit) $%
r_{ms}$ the condition $V_{,rr}(r)=0$ holds, condition from which we can
derive the value of $r_{ms}$ for a specified function $\nu (r)$.

After inserting Eqs. (\ref{Omega})-(\ref{L}) into the integral (\ref{F}),
and taking into account that $\sqrt{g}=r^2e^{(\nu+\lambda)/2}$ for $%
\theta=\pi/2$, we can compute the flux $F(r)$ over the whole disk surface
for each brane black hole geometry given by the metric functions $\nu(r)$
and $\lambda(r)$.

The accreting matter in the steady-state thin disk model is supposed to be
in thermodynamical equilibrium. Therefore the radiation emitted by the disk
surface can be considered as a perfect black body radiation, where the
energy flux is given by $F(r)=\sigma T^{4}(r)$ ($\sigma $ is the
Stefan-Boltzmann constant), and the luminosity $L\left( \omega \right) $ has
a black body spectrum \cite{To02}:
\begin{equation}
L\left( \omega \right) =4\pi d^{2}I\left( \omega \right) =\frac{4}{\pi }\cos
\gamma \omega ^{3}\int_{r_{i}}^{r_{f}}\frac{rdr}{\exp \left( \omega
/T\right) -1}.
\end{equation}

Here $d$ is the distance to the source, $I(\omega )$ is the Planck
distribution function, $\gamma $ is the disk inclination angle, and $r_{i}$
and $r_{f}$ indicate the position of the inner and outer edge of the disk,
respectively. We take $r_{i}=r_{ms}$ and $r_{f}\rightarrow \infty $, since
we expect the flux over the disk surface vanishes at $r\rightarrow \infty $
for any kind of brane black hole geometry.

The flux and the emission spectrum of the accretion disks around black holes
satisfy some simple scaling relations, with respect to the simple scaling
transformation of the radial coordinate, given by $r\rightarrow \bar{r}=r/M$,
where $M$ is the mass of the black hole. Generally, the metric tensor
coefficients are invariant with respect of this transformation, while the
specific energy, the angular momentum and the angular velocity transform as $%
\widetilde{E}\rightarrow \widetilde{E}$, $\widetilde{L}\rightarrow M\widetilde{L}$ and $%
\Omega \rightarrow \bar{\Omega}/M$, respectively. The flux scales as $F(r)\rightarrow F(%
\bar{r})/M^{4}$, giving the simple transformation law of the temperature as $%
T(r)\rightarrow T\left( \bar{r}\right) /M$. By also rescaling the frequency
of the emitted radiation as  $\omega \rightarrow \bar{\omega}=\omega /M$,
the luminosity of the disk is given by $L\left( \omega \right) \rightarrow
L\left( \bar{\omega}\right) /M$. On the other hand, the flux is proportional
to the accretion rate $\dot{M}_{0}$, and therefore an increase in the
accretion rate leads to a linear increase in the radiation emission flux
from the disc.

The efficiency $\epsilon $ with which the central object converts rest mass into
outgoing radiation is the other important physical parameter characterizing the properties of the accretion disks. The efficiency
is defined by the ratio of two rates measured at infinity: the rate of the
radiation of the energy of the photons escaping from the disk surface to infinity,
and the rate at which mass-energy is transported to the black hole. If all
the emitted photons can escape to infinity, the efficiency depends only on
the specific energy measured at the marginally stable orbit $r_{ms}$,
\begin{equation}
\epsilon = 1 - \widetilde{E}_{ms}\;.  \label{epsilon}
\end{equation}
For Schwarzschild black holes the efficiency is about 6\%, no matter if we
consider the photon capture by the black hole, or not. Ignoring the capture
of radiation by the black hole, $\epsilon$ is found to be 42\% for rapidly
rotating black holes, whereas the efficiency is 40\% with photon capture in
the Kerr potential.

\section{Thin disk accretion onto brane world black holes}

In the present Section we consider the accretion properties of several
classes of brane world black holes, which have been obtained by solving the
vacuum gravitational field equations for the metrics given by Eqs.~(\ref{metr1}%
) and ($\ref{rotmetr}$), respectively, where the metric functions $\nu $ and $\lambda $ depend only on $r$.
There are many black hole type solutions on the brane, and in the following
we analyze only four particular example, including three static black hole solutions, described by various metric
potentials $\nu (r)$ and $\lambda (r)$,  and the rotating generalization of the Kerr black hole. All the corresponding metric functions satisfy the gravitational field equations on the brane. The energy flux $F(r)$ and the disk emission
spectrum is obtained for supermassive brane world black holes, with a total mass of $2.5\times
10^{6}M_{\odot }$ and by using a mass accretion rate of $2\times
10^{-6}M_{\odot }$ yr$^{-1}$. The inclination angle $\gamma $ used for
the calculation of the spectra is set to $\cos \gamma =1$.

\subsection{The DMPR brane black hole}

The first brane black hole we consider is a solution of the vacuum field
equations, obtained by Dadhich, Maartens, Papadopoulos and Rezania in \cite%
{Da00}, which represent the simplest generalization of the Schwarzschild
solution of general relativity. We call this type of brane black hole as the
DMPR black hole. For this solution the metric tensor components are given by
\begin{equation}
e^{\nu }=e^{-\lambda }=1-\frac{2M}{r}+\frac{Q}{r^{2}} ,  \label{DMPR}
\end{equation}
where $Q$ is the so-called tidal charge parameter. In the limit $%
Q\rightarrow 0$ we recover the usual general relativistic case. The metric
is asymptotically flat, with $\lim _{r\to \infty}\exp{(\nu )}=\lim _{r\to
\infty}\exp{(\lambda )}=1$. There are two horizons, given by
\begin{equation}
r_{h}^{\pm}=M\pm \sqrt{M^{2}-Q}.
\end{equation}

Both horizons lie inside the Schwarzschild horizon $r_{s}=2M$, $0\leq
r_{h}^{-}\leq r_{h}^{+}\leq r_{s}$. In the brane world models there is also
the possibility of a negative $Q<0$, which leads to only one horizon $r_{h+}$
lying outside the Schwarzschild horizon,
\begin{equation}
r_{h+}=M+\sqrt{M^{2}+Q}>r_{s}.
\end{equation}

In this case the horizon has a greater area than its general relativistic
counterpart, so that bulk effects act to increase the entropy and decrease
the temperature, and to strengthen the gravitational field outside the black
hole.

\begin{figure}[tbp]
\includegraphics[width=8.7cm]{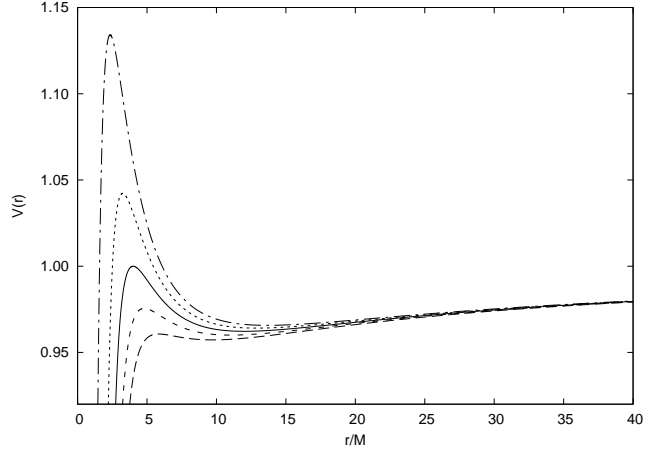}
\caption{The effective potential of a DMPR black hole with total mass $M$
and tidal charge $Q$ for the specific angular momentum $\widetilde{L}%
=4M $. The solid line is the effective potential for a Schwarzschild black
hole with the same total mass. The values of $Q$ are $Q=-M^2$
(long dashed line), $Q=-0.5M^{2}$ (short dashed line), $Q=0.5M^{2}$ (dotted
line) and $Q=M^{2}$ (dot-dashed line), respectively.}
\label{DMPR_V}
\end{figure}

DMPR brane black holes are characterized by the metric functions (\ref{DMPR}%
). If we insert $e^\nu$ from these equations into Eq.~(\ref{V2}), we obtain
the effective potential $V(r)$ for this type of black hole for any
particle with a specific angular momentum $%
\widetilde{L}$, orbiting around the black hole, as a function of
the total mass $M$, and of the tidal charge $Q$ of the black hole. In Fig.~\ref{DMPR_V} we present the radial profile of $V(r)$
for $M/\widetilde{L}=4$, and different values of the tidal charge $Q$,
running between $-M^2$ and $M^2$. For comparison we have also plotted  the Schwarzschild
potential, corresponding to $Q=0$.

\begin{figure}[tbp]
\includegraphics[width=8.15cm]{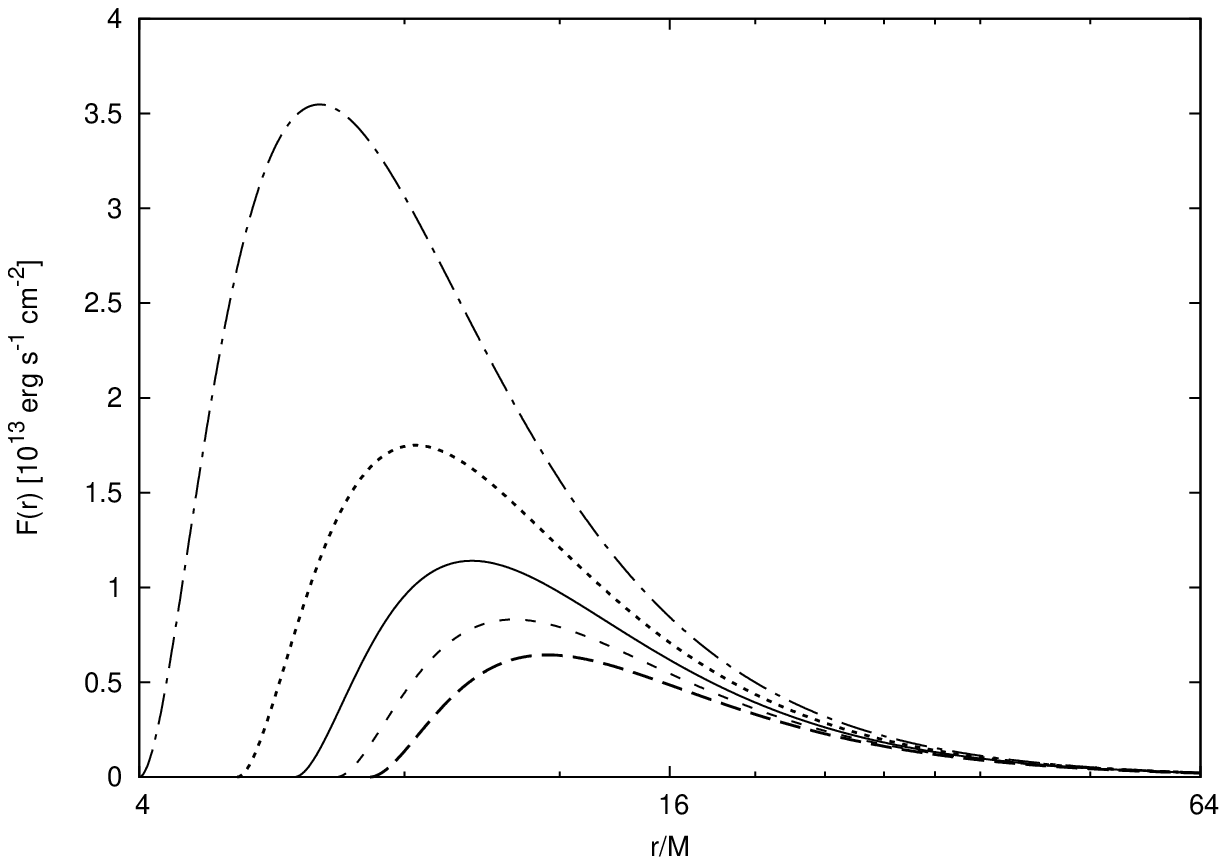}
\includegraphics[width=8.15cm]{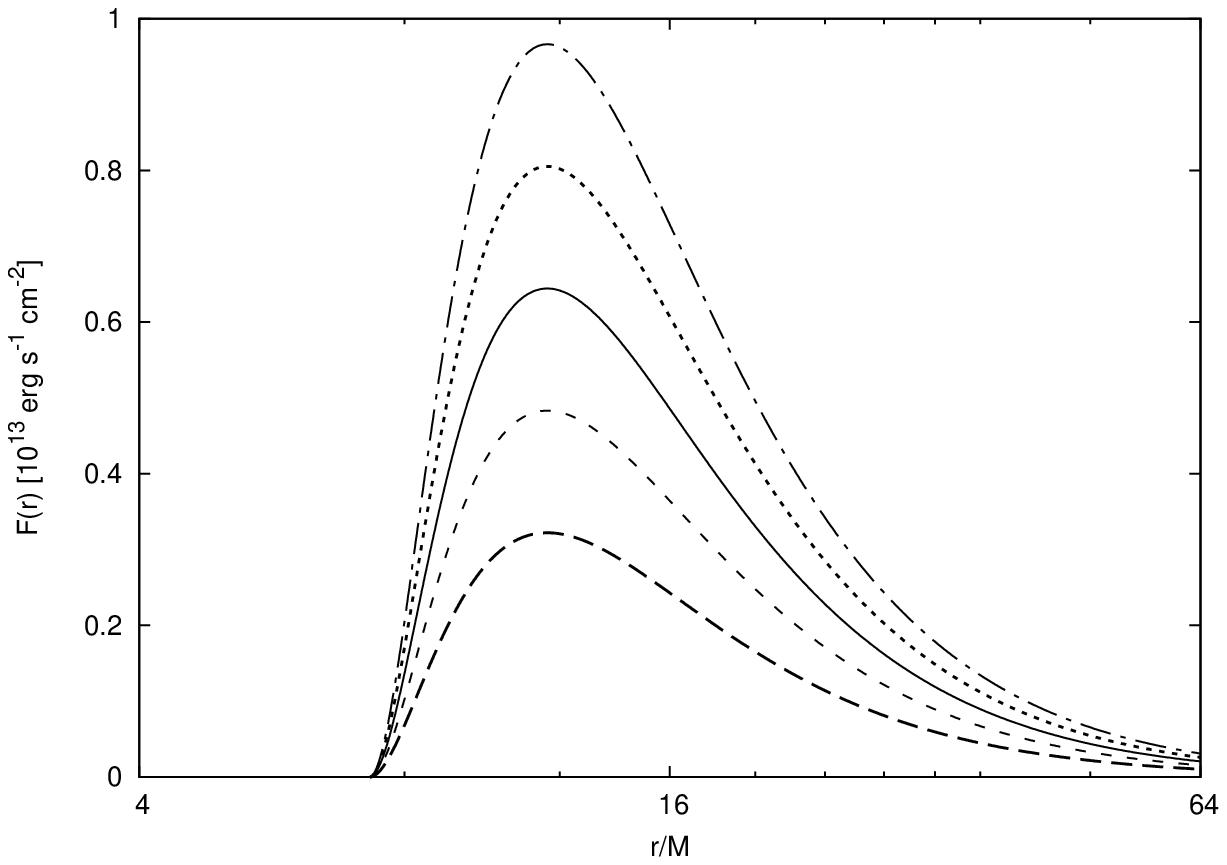}
\caption{Figure on the left: The time-averaged flux radiated by the accretion disk around a DMPR
black hole for different
values of the tidal charge parameter $Q$: $Q=-M^2$ (long dashed line), $Q=-0.5M^{2}$ (short dashed line), $%
Q=0.5M^{2}$ (dotted line) and $Q=M^{2}$ (dot-dashed line), respectively. The flux for a Schwarzschild black
hole is plotted with a solid line. The
mass accretion rate is $2\times 10^{-6} M_{\odot}$yr$^{-1}$. Figure on the right: The time-averaged flux radiated by the accretion disk around a DMPR black hole and tidal charge $Q=-M^2$ for different values of the mass accretion rate $\dot{M}_0$: $1.0\times10^{-6} M_{\odot}$yr$^{-1}$ (long dashed line), $1.5\times10^{-6} M_{\odot}$yr$^{-1}$ (short dashed line), $=2.0\times10^{-6} M_{\odot}$yr$^{-1}$ (solid line), $2.5\times 10^{-6} M_{\odot}$yr$^{-1}$ (dotted line), and $3.0\times 10^{-6} M_{\odot}$yr$^{-1}$ (dot-dashed line), respectively.  The total mass of the black hole is $2.5\times10^6 M_{\odot}$  }
\label{DMPR_F}
\end{figure}

By increasing $Q$ from zero to $M^2$ we also increase the potential barrier,
as compared to the Schwarzschild case, whereas negative tidal charges lowers
the barrier, as expected for the potential of the Reissner-Nordstr\"om type
black holes. The variation of $Q$ also modifies the position of the
marginally stable orbit, as shown by the shift of the cut-off in the left
hand side of the radial profiles of the photon flux $F(r)$ over the disk
surface, which is presented in the left plot in Fig.~\ref{DMPR_F}.

For positive increasing tidal charges, $r_{ms}$ takes lower values, and the
negative decreasing charges lead to the increase of $r_{ms}$. The position of the
maximum of the radiation flux is changed similarly due to the variations of $%
r_{ms}$. By increasing the value of $Q$ from $-M^2$ to $M^2$, we shift the
radius of the maximal flux to lower and lower values, approaching the
marginally stable orbit. The vertical shift in the effective potential
modifies the amplitude of the flux profile as well: higher values of the
potential for $Q=0.5M^2$ and $Q=M^2$ involve a higher specific energy of the
orbiting particles, and therefore  higher flux values. For $%
Q=-0.5M^2$ and $Q=-M^2$, the bound state of the particles has a lower
energy, and a smaller amount of energy is radiated away from the disk
surface, which decreases the peak in the flux profile. In the right plot of Fig.~\ref{DMPR_F} we present the variation of the flux for different values of the mass accretion rate.

\begin{figure}[tbp]
\includegraphics[width=8.15cm]{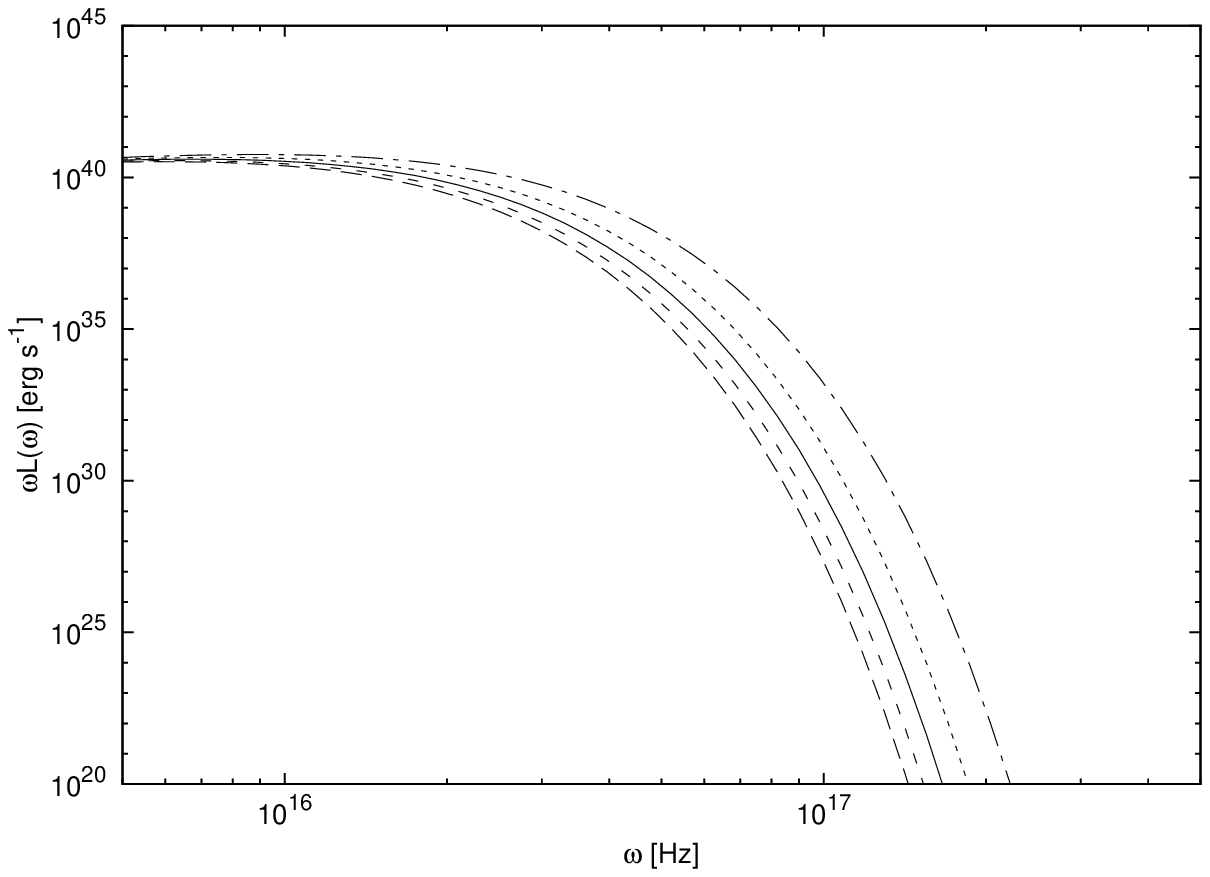}
\includegraphics[width=8.15cm]{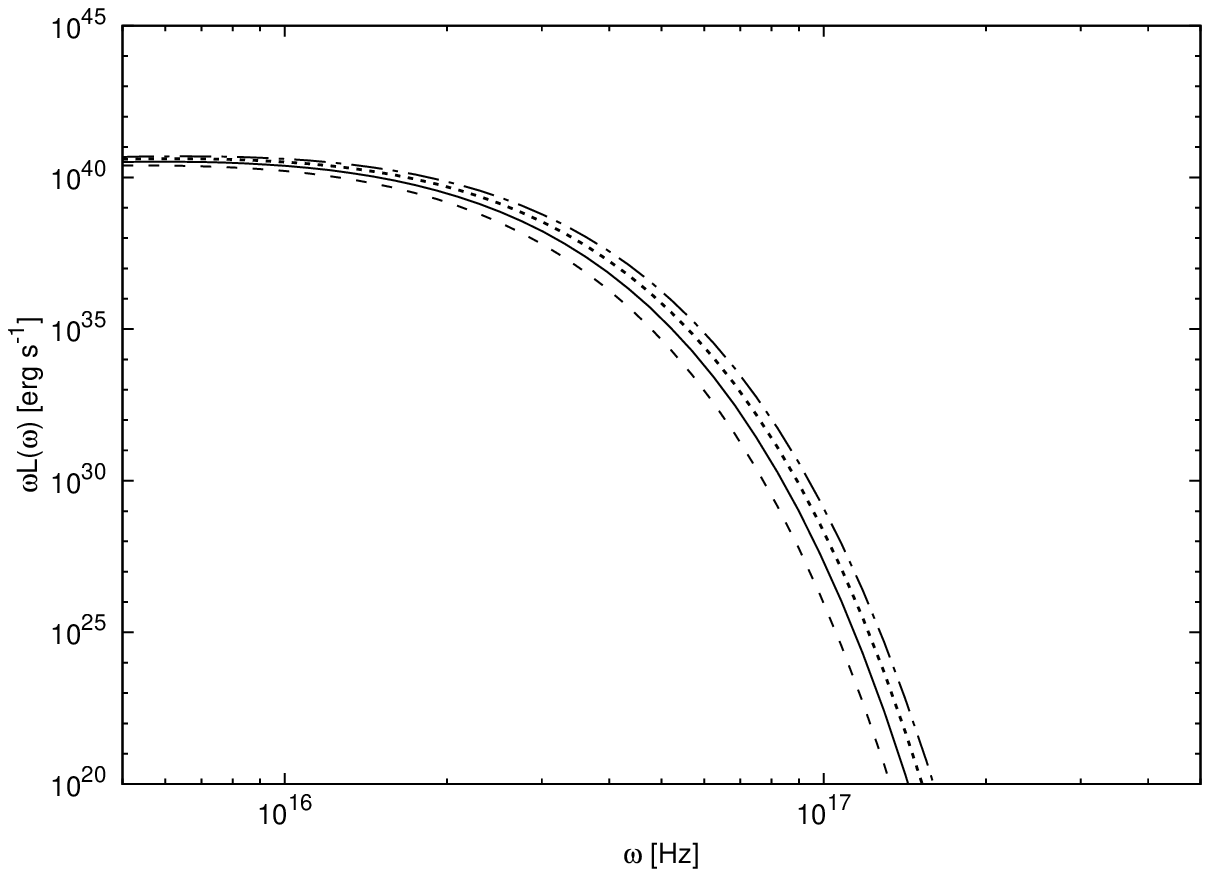}
\caption{Figure on the left: The emission spectrum of the accretion disk around a DMPR black
hole for different
values of the tidal charge parameter $Q$: $Q=-M^2$ (long dashed line), $Q=-0.5M^{2}$ (short dashed line), $%
Q=0.5M^{2}$ (dotted line) and $Q=M^{2}$ (dot-dashed line), respectively.  The solid line represents the disk
spectrum for a Schwarzschild black hole. The
mass accretion rate is $2\times 10^{-6} M_{\odot}$yr$^{-1}$. Figure on the right: The emission spectrum of the accretion disk around a DMPR black hole with  tidal charge $Q=-M^2$ for different values of the mass accretion rate $\dot{M}_0$: $1.0\times10^{-6} M_{\odot}$yr$^{-1}$ (long dashed line), $1.5\times10^{-6} M_{\odot}$yr$^{-1}$ (short dashed line), $2.0\times10^{-6} M_{\odot}$yr$^{-1}$ (solid line),
$2.5\times 10^{-6} M_{\odot}$yr$^{-1}$ (dotted line), and $3.0\times 10^{-6} M_{\odot}$yr$^{-1}$ (dot-dashed line), respectively. The total mass of the black hole is $2.5\times10^6 M_{\odot}$ }

\label{DMPR_L}
\end{figure}

These effects also cause a shift of the cut-off frequency of the black body
emission spectrum of the accretion disk, as we can see in the left plot of Fig.~\ref{DMPR_L}.
The increasing positive tidal charges shift the cut-off to higher
frequencies, which produces a harder spectrum, whereas the negative
decreasing values of $Q$ soften the disk spectrum by lowering the cut-off
frequencies.

The variation of the emission spectrum for different values of the mass accretion rate is presented in the right plot in Fig.~\ref{DMPR_L}. We also present the conversion efficiency $\epsilon$ of the  mass accreted by the DMPR black hole
into radiation measured at infinity, which is given by Eq.~(\ref{epsilon}).
The value of $\epsilon$ measures the efficiency of energy generating
mechanism by mass accretion. The amount of energy released by the matter leaving
the marginally stable orbit, and falling into the black hole, is the binding
energy $\widetilde{E}_{ms}$ of the black hole potential. For different tidal
charges the values of $\widetilde{E}_{ms}$, together with the radii of the
marginally stable orbits, are given in Table~\ref{DMPR_eps}, where the
figures corresponding to the Schwarzschild black hole corresponds to $Q=0$. As we
increase $Q$ from $-M^2$ to $M^2$, $r_{ms}$ is decreasing from values
greater than the radius of the marginally stable orbit for the Schwarzschild
geometry to lower ones. The efficiency has an opposite trend: for negative
tidal charges it has smaller values than in the case of the Schwarzschild
black holes, and it exceeds the latter for $Q>0$.

\begin{table}[tbp]
\begin{center}
\begin{tabular}{|c|c|c|}
\hline
$Q$ [$M^2$] & $r_{ms}$ [$M$] & $\epsilon$ \\ \hline
-1 & 7.3100 & 0.0476 \\ \hline
-0.5 & 6.6949 & 0.0517 \\ \hline
0 & 6.0000 & 0.0572 \\ \hline
0.5 & 5.1695 & 0.0655 \\ \hline
1 & 4.0019 & 0.0814 \\ \hline
\end{tabular}%
\end{center}
\caption{The marginally stable orbit and the efficiency for different DMPR
black hole geometries. The case $Q=0$ corresponds to the standard
general relativistic Schwarzschild black hole.}
\label{DMPR_eps}
\end{table}

\subsection{The CFM brane black hole}

Two families of analytic solutions in the brane world model, parameterized
by the ADM mass and the PPN parameters $\beta $ and $\gamma $, and which
reduce to the Schwarzschild black hole for $\beta =1$, have been found by
Casadio, Fabbri and Mazzacurati in \cite{cfm02}. We call the corresponding
brane black holes as the CFM black holes.

The first class of solutions is given by
\begin{equation}
e^{\nu }=1-\frac{2M}{r},  \label{cfm020}
\end{equation}%
and
\begin{equation}
e^{\lambda }=\frac{1-\frac{3M}{r}}{\left( 1-\frac{2M}{r}\right) \left[ 1-%
\frac{3M}{2r}\left( 1+\frac{4}{9}\eta \right) \right] },
\end{equation}%
respectively, where $\eta =\gamma -1=2\left( \beta -1\right) $. As in the
Schwarzschild case the event horizon is located at $r=r_{h}=2M$. The
solution is asymptotically flat, that is $\lim_{r\rightarrow \infty }e^{\nu
}\equiv e^{\nu _{\infty }}\equiv \lim_{r\rightarrow \infty }e^{\lambda
}\equiv e^{\lambda _{\infty }}=1$.

The second class of solutions corresponding to brane world black holes
obtained in \cite{cfm02} has the metric tensor components given by
\begin{equation}  \label{cfm021}
e^{\nu }=\left[ \frac{\eta +\sqrt{1-\frac{2M}{r}\left( 1+\eta \right) }}{%
1+\eta }\right] ^{2},
\end{equation}
and
\begin{equation}  \label{cfm022}
e^{\lambda }=\left[ 1-\frac{2M}{r}\left( 1+\eta \right) \right] ^{-1},
\end{equation}
respectively. The metric is asymptotically flat. In the case $\eta >0$, the
only singularity in the metric is at $r=r_{0}=2M\left( 1+\eta \right) $,
where all the curvature invariants are regular. $r=r_{0}$ is a turning point
for all physical curves. For $\eta <0$ the metric is singular at $%
r=r_{h}=2M/\left( 1-\eta \right) $ and at $r_{0}$, with $r_{h}>r_{0}$. $%
r_{h} $ defines the event horizon.

The first and second class of the CFM brane black holes are characterized by
the metric potentials (\ref{cfm020}) and (\ref{cfm021}). Since for the first
class of solutions the metric function $e^{\nu}$ coincides with the one of
the Schwarzschild black hole, their effective potentials $V(r)$ are the same
for equal total masses and fixed $\tilde L$. As a consequence, the specific
energy, specific angular momentum and angular velocity of the particles
orbiting around the first class of CFM black holes are equal to those of the
particles moving at the same Keplerian orbit in the Schwarzschild potential.

\begin{figure}[tbp]
\includegraphics[width=8.15cm]{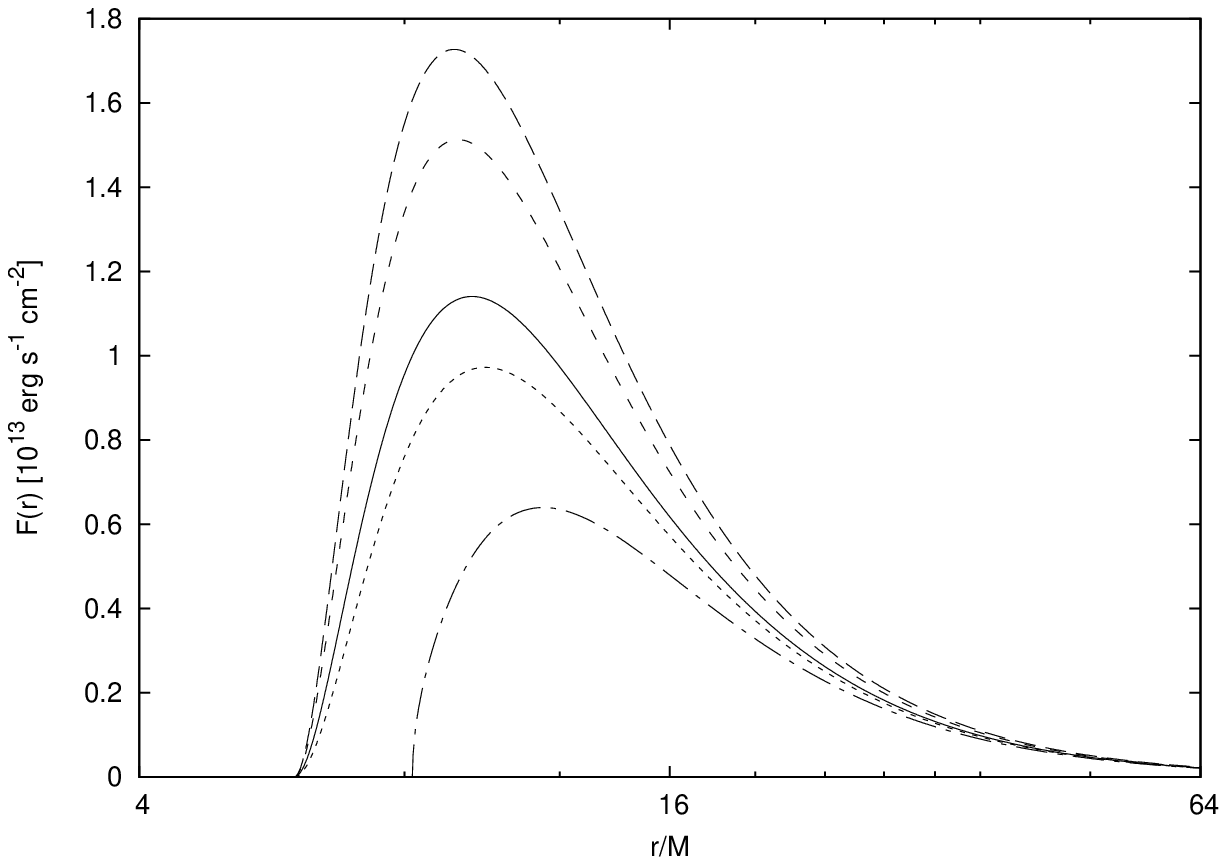}
\includegraphics[width=8.15cm]{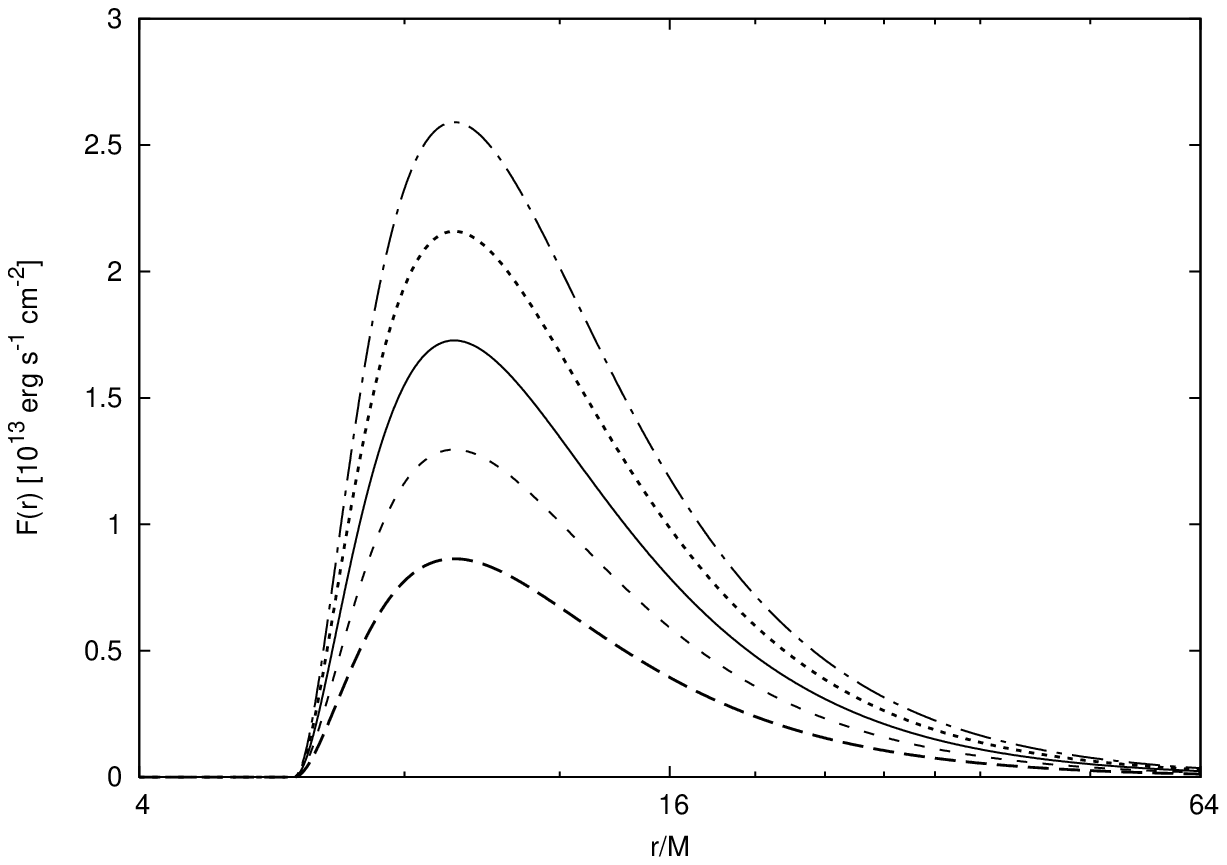}
\caption{Figure on the left: The time-averaged flux radiated by an accretion disk around a first
class CFM black holes
for various values of $\eta$: $\eta=-10$ (long dashed line), $\eta=-5$
(short dashed line), $\eta=5$ (dotted line) and $\eta=10$
(dot-dashed line).  The flux for a Schwarzschild black hole
 is plotted with a solid line. The mass accretion rate is $2\times
10^{-6} M_{\odot}$ yr$^{-1}$. Figure on the right: The time-averaged flux radiated by the accretion disk around a first class CFM
black hole for $\eta =-10$ and for different values of the mass accretion rate $\dot{M}_0$: $1.0\times10^{-6} M_{\odot}$yr$^{-1}$ (long dashed line), $1.5\times10^{-6} M_{\odot}$yr$^{-1}$ (short dashed line), $2.0\times10^{-6} M_{\odot}$yr$^{-1}$ (solid line),
$2.5\times 10^{-6} M_{\odot}$yr$^{-1}$ (dotted line), and $3.0\times 10^{-6} M_{\odot}$yr$^{-1}$ (dot-dashed line), respectively. The total mass of the black hole is $2.5\times 10^{6} M_{\odot}$.}
\label{CMF1_F}
\end{figure}

However, the radiation flux from the accretion disk shows considerable
differences as compared to the Schwarzschild case when the parameter $\eta$ is varied. This behavior, shown in
the left plot of Fig.~\ref{CMF1_F}, is due to the fact that the proper volume used in the
calculation of any integral in this spacetime depends on both the metric
functions $\nu(r)$ and $\lambda(r)$, and in the flux integral given by Eq.~(\ref{F}) we have
\begin{equation}
\sqrt{-g}=r\left[\frac{1-\frac{3M}{r}}{1-\frac{3M}{2r}(1+\frac{4}{9}\eta)}%
\right]^{1/2}.  \label{sqrtg}
\end{equation}
Since the left hand side of Eq.
(\ref{sqrtg}) is decreasing for negative values of $\eta$, we obtain higher and
higher flux values by decreasing this metric parameter for $\eta<0$. We have
the opposite effect for $\eta>0$, where the increase of this parameter
causes the proper volume to also increase, and, in turn, the amplitudes of
the flux profile to decrease. Because the effective potential does not
depend on $\eta $, there is no variation in the radius of the marginally
stable orbit. However, a cut-off appears in the left hand side of the flux
profiles for $\eta>0$, where the denominator in Eq.~(\ref{sqrtg}) becomes
negative. This gives the criterion
\begin{equation}
\frac{r}{M}<\frac{3}{2}\left(1+\frac{4M}{9}\right)\;.
\end{equation}
As one can see in left plot of Fig.~\ref{CMF1_F}, for $\eta=5$ this criterion is true for any radii
greater than $r_{ms}$, but we obtain a cut-off in the left hand
side of the flux profile by setting $\eta$ equal to $\eta =10$.

Although the position of the marginally stable orbit does not change for
different values of $\eta$, the maximum of the radial distribution of the
flux is located at higher and higher radii as we increase $\eta $. The right plot in Fig.~\ref{CMF1_F} presents the variation of the flux for a fixed mass and parameter $\eta $, and for different values of the accretion rate.

\begin{figure}[tbp]
\includegraphics[width=8.15cm]{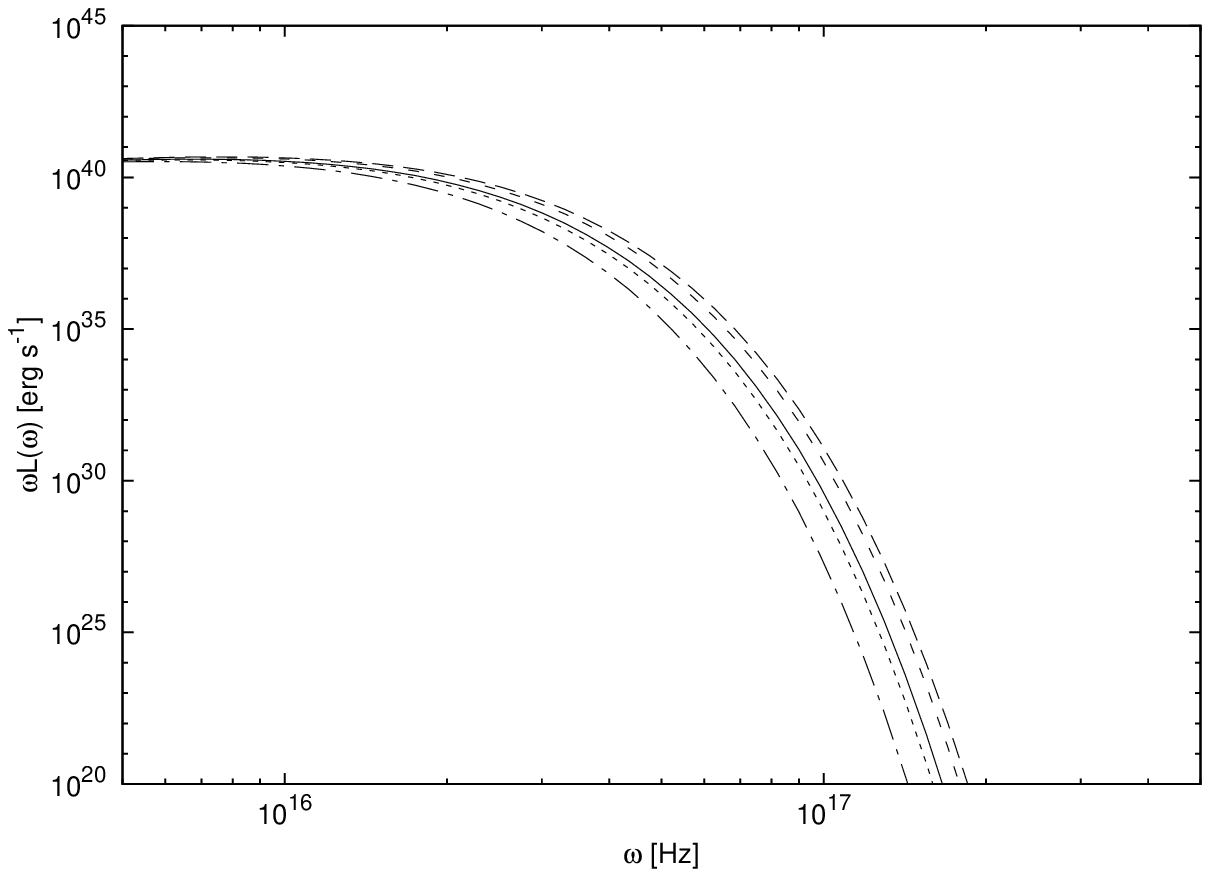}
\includegraphics[width=8.15cm]{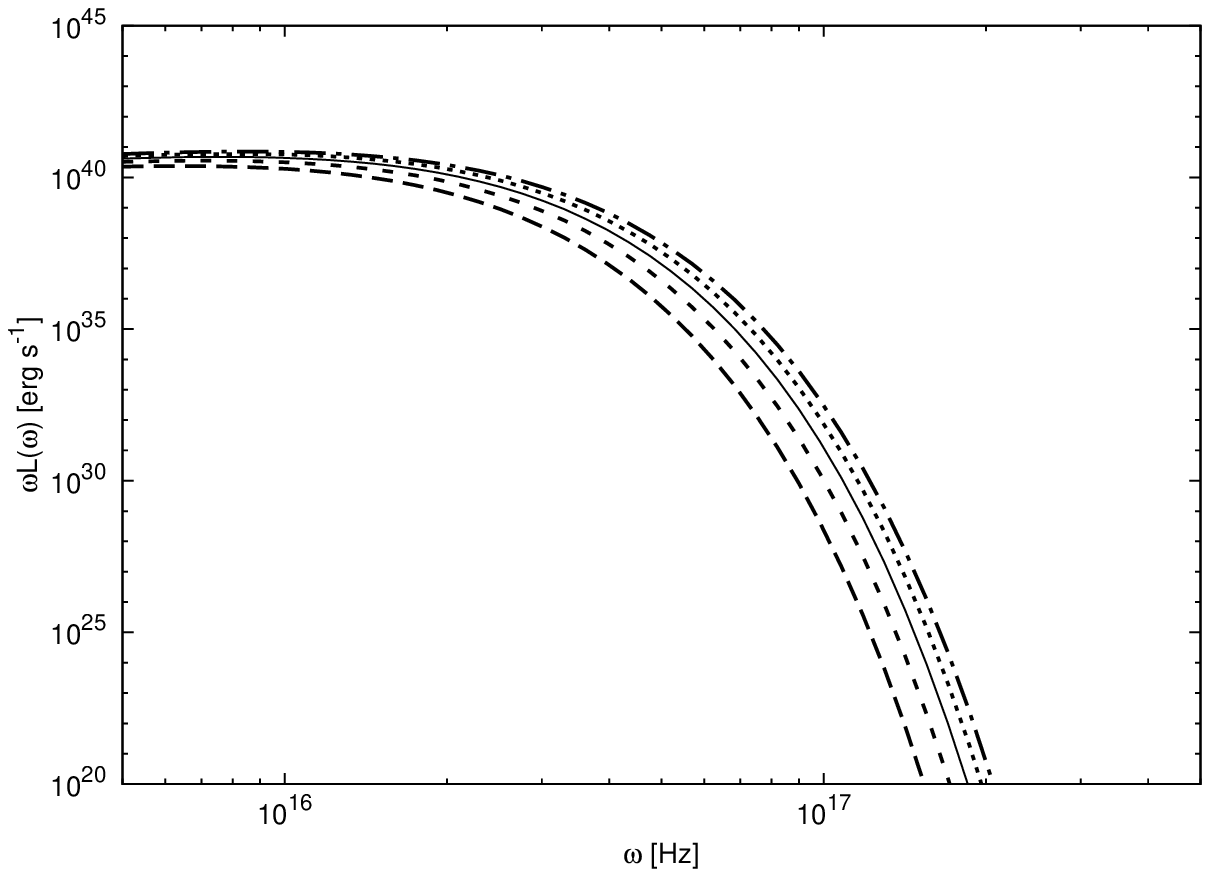}
\caption{Figure on the left: The emission spectrum of the accretion disk around a first class CFM black hole
for different values of $\eta $: $\eta =-10$ (long dashed line), $%
\eta =-5$ (short dashed line), $\eta=5$ (dotted line) and $%
\eta =10$ (dot-dashed line). The spectrum for a Schwarzschild black
hole is plotted with a solid line. The mass
accretion rate is $2\times 10^{-6} M_{\odot}$ yr$^{-1}$. Figure on the right: The emission spectrum of the accretion disk around a first class CFM
black hole for $\eta =-10$ and for different values of the mass accretion rate $\dot{M}_0$: $=1.0\times10^{-6} M_{\odot}$yr$^{-1}$ (long dashed line), $1.5\times10^{-6} M_{\odot}$yr$^{-1}$ (short dashed line), $=2.0\times10^{-6} M_{\odot}$yr$^{-1}$ (solid line),
$2.5\times 10^{-6} M_{\odot}$yr$^{-1}$ (dotted line), and $3.0\times 10^{-6} M_{\odot}$yr$^{-1}$ (dot-dashed line), respectively. The total mass of the black hole is $2.5\times 10^{6} M_{\odot}$. }
\label{CMF1_L}
\end{figure}

The variation in the shape and the amplitude of the flux profile for different
values of $\eta$ has a clear effect of shifting the cut-off frequency in the
disk spectra. The left plot of Fig.~\ref{CMF1_L} shows that the cut-off value shifts to
higher frequencies when $\eta $ is negative, hardening the spectra. For $%
\eta>0$ the disk spectrum becomes softer, with lower cut-off frequencies. The effect of the variation of the mass accretion rate on the emission spectra for a fixed mass and $\eta $ is presented in the right plot of Fig.~\ref{CMF1_L}.\texttt{}

 Since the effective potential of the CFM black holes of the first class is
the same as the one of the Schwarzschild geometry, their conversion
efficiencies $\epsilon$ of the mass accreted by the black hole into radiation measured at
infinity are also the same.

\begin{figure}[tbp]
\includegraphics[width=8.7cm]{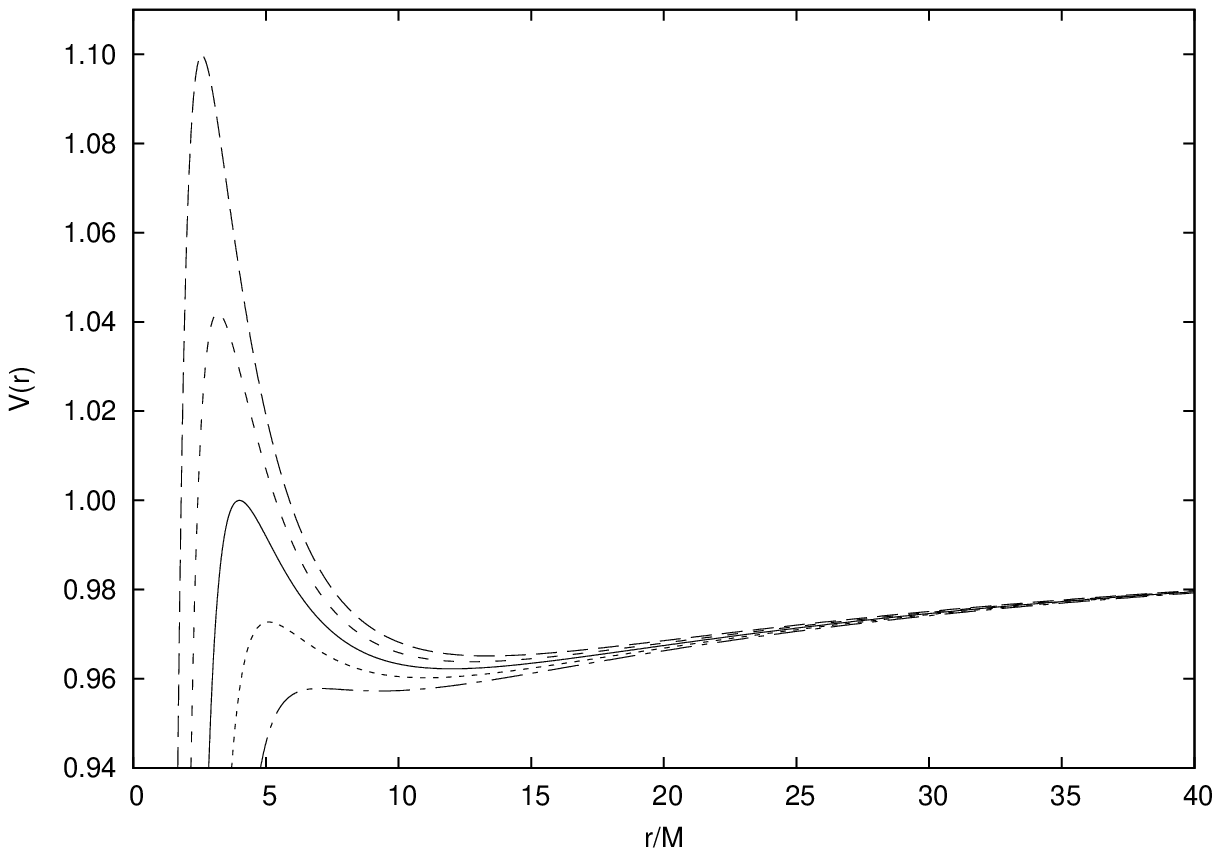}
\caption{The effective potential of the thin accretion disk around a CFM
black hole of the second class with a total mass of $M$ and $\widetilde{L}%
=4M $ for different values of the parameter $\eta $. The potential $%
V(r)$ corresponding to the Schwarzschild black hole is plotted with a solid
line. The values of $\eta$ are $\eta =-0.8$ (long dashed
line), $\eta =-0.4$ (short dashed line), $\eta =0.4$ (dotted
line) and $\eta = 0.8$ (dot-dashed line), respectively.}
\label{CMF2_V}
\end{figure}

If we consider the second class of the CFM black holes, given by Eqs. (\ref%
{cfm021}) and (\ref{cfm022}), respectively, the variation of the parameter $%
\eta $ causes similar effects in the behavior of the photon flux emitted by
the accretion disk and its spectrum. However, the metric function $\nu(r)$
now differs from the Schwarzschild black hole case, and we obtain
different effective potentials for different values of the $\eta $. The
radial profiles of $V(r)$ are shown in Fig.~\ref{CMF2_V}, where $\eta$ is
set to values between $-0.8$ and $0.8$.

\begin{figure}[tbp]
\includegraphics[width=8.15cm]{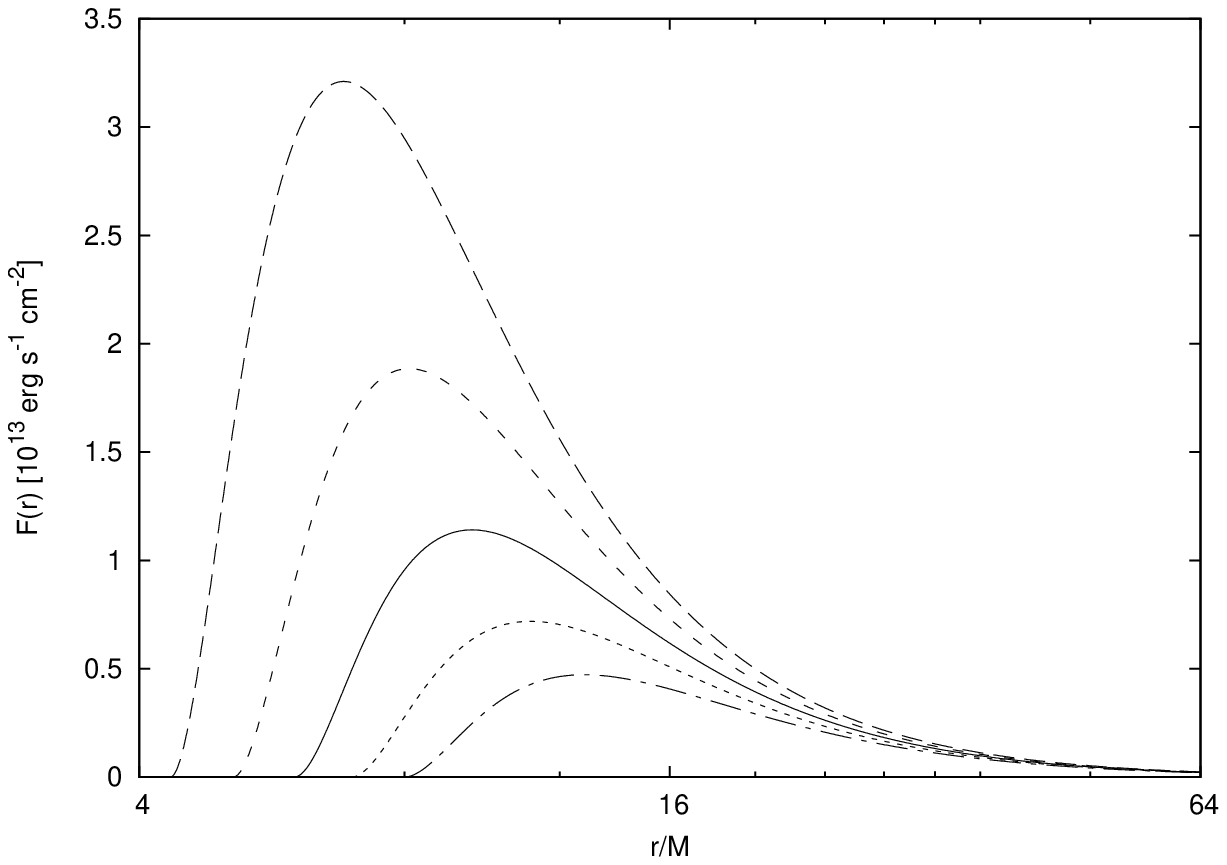}
\includegraphics[width=8.15cm]{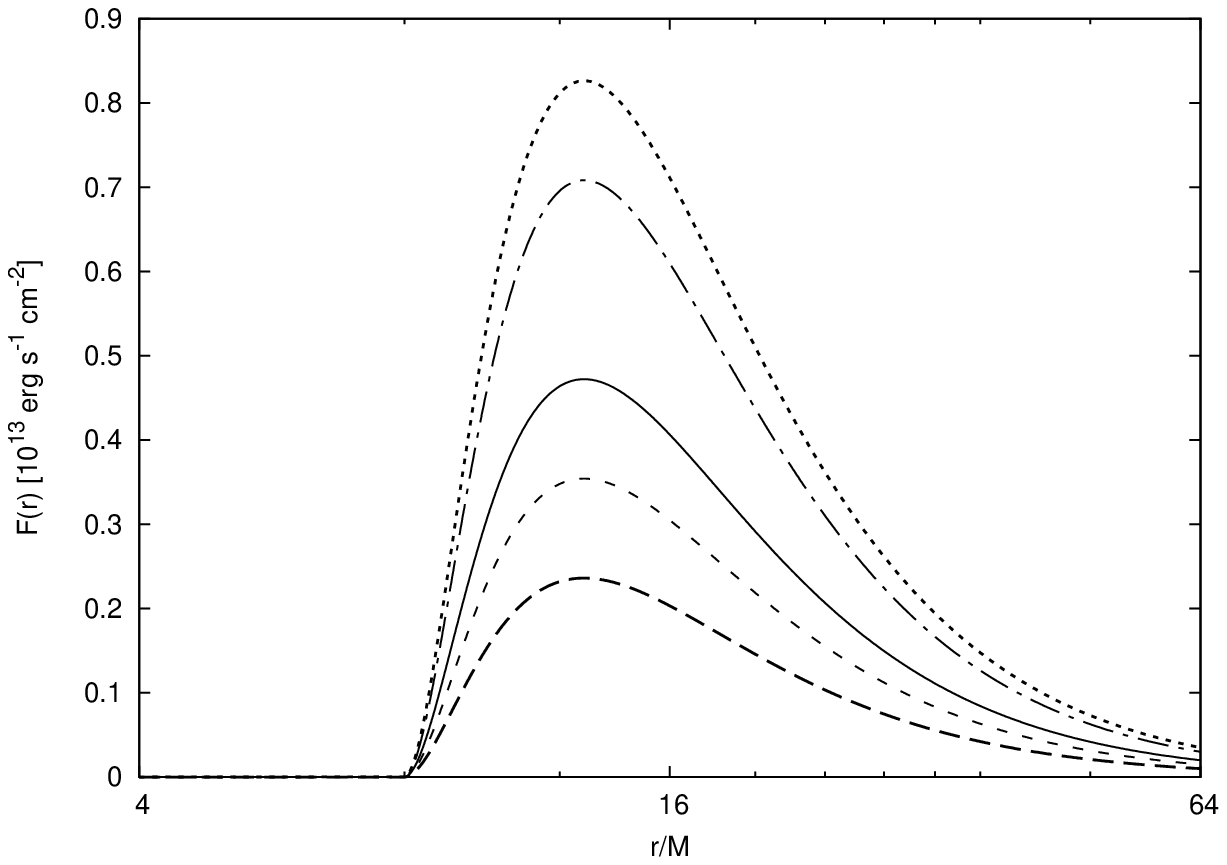}
\caption{Figure on the left: The time-averaged flux radiated by an accretion disk around a second class CFM
black hole  for different values of $\eta$: $\eta =-0.8$
(long dashed line), $\eta =-0.4$ (short dashed line), $\eta %
=0.4$ (dotted line) and $\eta =0.8$ (dot-dashed line). The flux for
a Schwarzschild black hole is plotted with a solid
line. The mass accretion rate is $2\times 10^{-6} M_{\odot}$ yr$^{-1}$. Figure on the right: The time-averaged flux radiated by the accretion disk around a second class CFM black hole with the parameter $\eta =0.8$ for different values of the mass accretion rate $\dot{M}_0$: $1.0\times10^{-6} M_{\odot}$yr$^{-1}$ (long dashed line), $1.5\times10^{-6} M_{\odot}$yr$^{-1}$ (short dashed line), $2.0\times10^{-6} M_{\odot}$yr$^{-1}$ (solid line),
$2.5\times 10^{-6} M_{\odot}$yr$^{-1}$ (dotted line), and $3.0\times 10^{-6} M_{\odot}$yr$^{-1}$ (dot-dashed line), respectively. The total mass of the black hole is $2.5\times 10^{6}
M_{\odot}$. }
\label{CMF2_F}
\end{figure}

By decreasing the values of $\eta $ from zero to small negative values, we
can increase the potential barrier around the black hole, and decrease the
radius of the marginally stable orbit. The minimum of the effective
potential is also increased. Any increment in $\eta $ from zero to small
positive values causes the opposite effects for $V(r)$. Then the parameters $%
\eta<0$ increase the flux values and shift $r_{ms}$ to lower radii, while
negative values of $\eta$ give lower fluxes with a cut-off at higher values
of $r_{ms}$, as shown in the left plot of Fig.~\ref{CMF2_F}.

\begin{figure}[tbp]
\includegraphics[width=8.15cm]{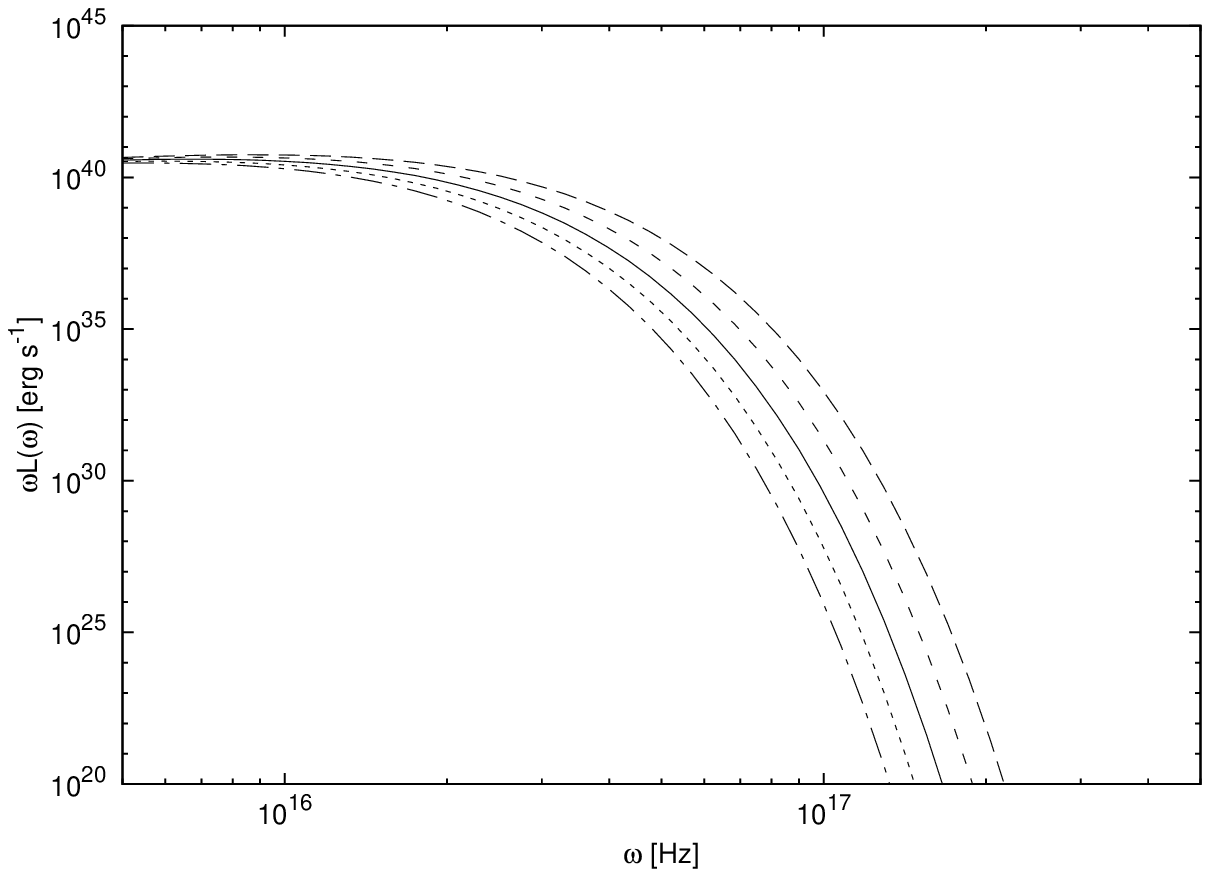}
\includegraphics[width=8.15cm]{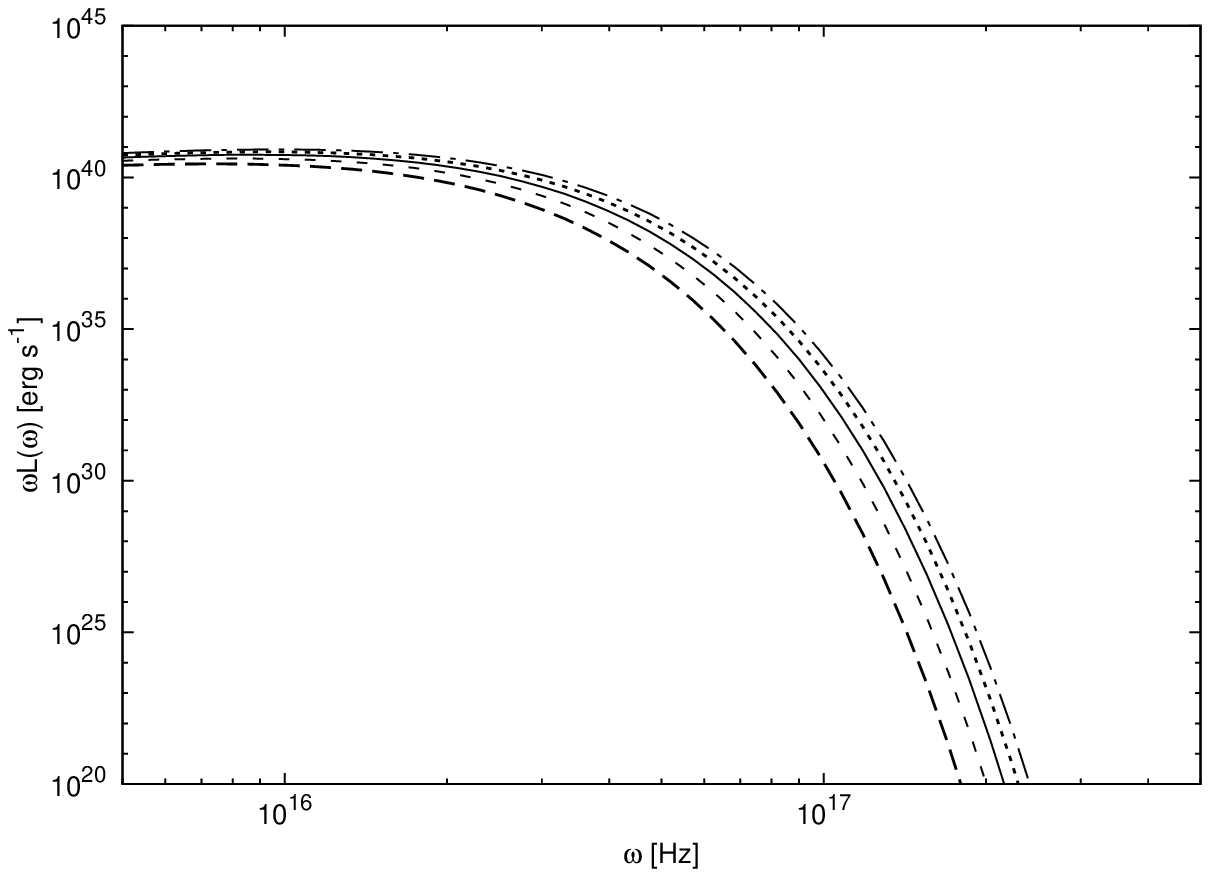}
\caption{Figure on the left: The emission spectrum of the thin accretion disk around a second class CFM black
hole for different values of $\eta$: $\eta =-0.8$
(long dashed line), $\eta =-0.4$ (short dashed line), $\eta %
=0.4$ (dotted line), and $\eta =0.8$ (dot-dashed line). The spectrum
for a Schwarzschild black hole  is plotted with a
solid line. The mass accretion rate is $2\times 10^{-6} M_{\odot}$ yr$%
^{-1}$. Figure on the right: The emission spectrum of the accretion disk around a second class CFM
black hole  with parameter $\eta =10$  for different values of the mass accretion rate $\dot{M}_0$: $1.0\times10^{-6} M_{\odot}$yr$^{-1}$ (long dashed line), $1.5\times10^{-6} M_{\odot}$yr$^{-1}$ (short dashed line), $2.0\times10^{-6} M_{\odot}$yr$^{-1}$ (solid line),
$2.5\times 10^{-6} M_{\odot}$yr$^{-1}$ (dotted line), and $3.0\times 10^{-6} M_{\odot}$yr$^{-1}$ (dot-dashed line), respectively. The total mass of the black hole is $2.5\times 10^{6}
M_{\odot}$. }
\label{CMF2_L}
\end{figure}

The variation of the numerical value of the radial coordinate $r$ at the
position where the radiation flux takes its maximum value follows the
tendency of $r_{ms}$: with increasing $\eta $ we obtain higher and higher
orbits for $F_{max}$. These effects result in similar shifts in the disk
spectrum as in the case of the CFM black holes of the first class: in the left plot of Fig.~%
\ref{CMF2_L} negative values of $\eta$ harden the spectrum, by shifting the
frequency cut-off to higher frequencies, whereas the spectra become softer,
with lower cut-off values, for increasing positive values of the metric
parameter.

The right plots of Figs.~\ref{CMF2_F} and \ref{CMF2_L} show the variation of the flux and emission spectra for a fixed mass and $\eta $ and for different values of the accretion rate.

We present the marginally stable orbits $r_{ms}$ and the conversion efficiency $\epsilon$ of the
matter accreted by the CFM black holes of
the second class into radiation measured at infinity in Table~\ref{CMF2_eps}. With the increasing values of the
parameter $\eta$, the radius of the marginally stable orbit is also
increasing, as already seen in the flux diagram, while $\epsilon$ drops down
from values greater than $6$ to lower ones.

\begin{table}[tbp]
\begin{center}
\begin{tabular}{|c|c|c|}
\hline
$\eta$ & $r_{ms}$ [$M$] & $\epsilon$ \\ \hline
-0.8 & 4.3445 & 0.0795 \\ \hline
-0.4 & 5.1165 & 0.0657 \\ \hline
0 & 6.0000 & 0.0572 \\ \hline
0.4 & 6.9722 & 0.0501 \\ \hline
0.8 & 8.0091 & 0.0444 \\ \hline
\end{tabular}%
\end{center}
\caption{The marginally stable orbit and the efficiency for CFM black holes
of the second kind. The case $\eta=0$ corresponds to the standard
general relativistic Schwarzschild black hole.}
\label{CMF2_eps}
\end{table}

\subsection{The BMD brane world black hole}

Several classes of brane world black hole solutions have been obtained by
Bronnikov, Melnikov and Dehnen in \cite{BMD03} (for short the BMD black
holes). In the following we analyze the accretion properties of a particular
class of these models, with metric given by
\begin{equation}
e^{\nu }=\left( 1-\frac{2M}{r}\right) ^{2/s},e^{\lambda }=\left( 1-\frac{2M}{%
r}\right) ^{-2},  \label{bmd03}
\end{equation}%
where $s\in $N. The metric is asymptotically flat, and at $r=r_{h}=2M$ these
solutions have a double horizon.

\begin{figure}[tbp]
\includegraphics[width=8.7cm]{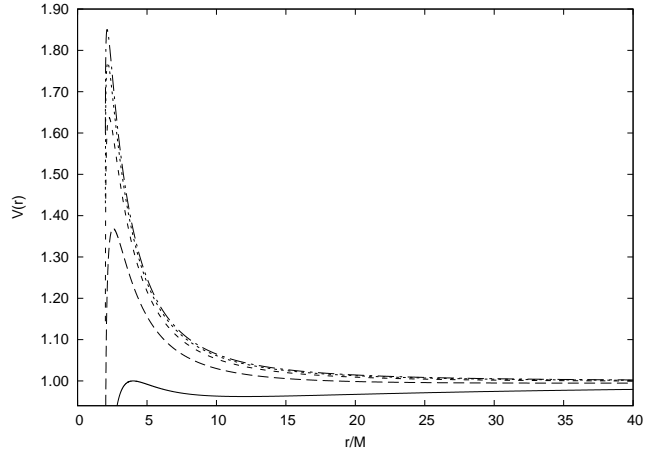}
\caption{The effective potential for BMD black holes of a fixed total mass $%
M $ for $\widetilde{L}=4M$ and different values of $s$. The solid line is
the effective potential for a Schwarzschild black hole with the same total
mass $M$. The parameter $s$ is set to $s=5$ (long dashed line), $s=10$
(short dashed line), $s=15$ (dotted line) and $s=20$ (dot-dashed line).}
\label{BMD_V}
\end{figure}

Eqs. (\ref{bmd03}) determine the geometry of the BMD brane black holes,
which have the effective potential plotted in Fig.~\ref{BMD_V}. In the
figure we have plotted $V(r)$ for values of $s$ between $s=5$ and $s=20$.

\begin{figure}[tbp]
\includegraphics[width=8.15cm]{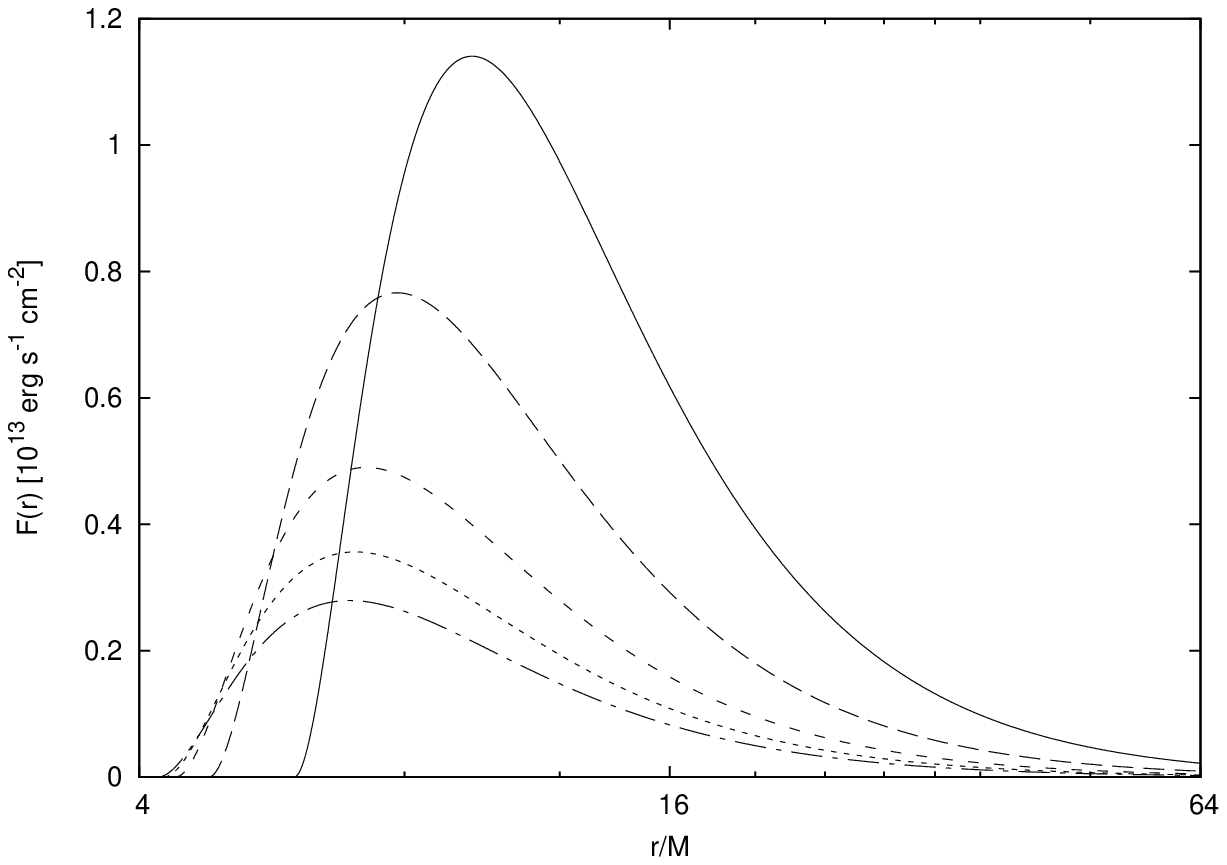}
\includegraphics[width=8.15cm]{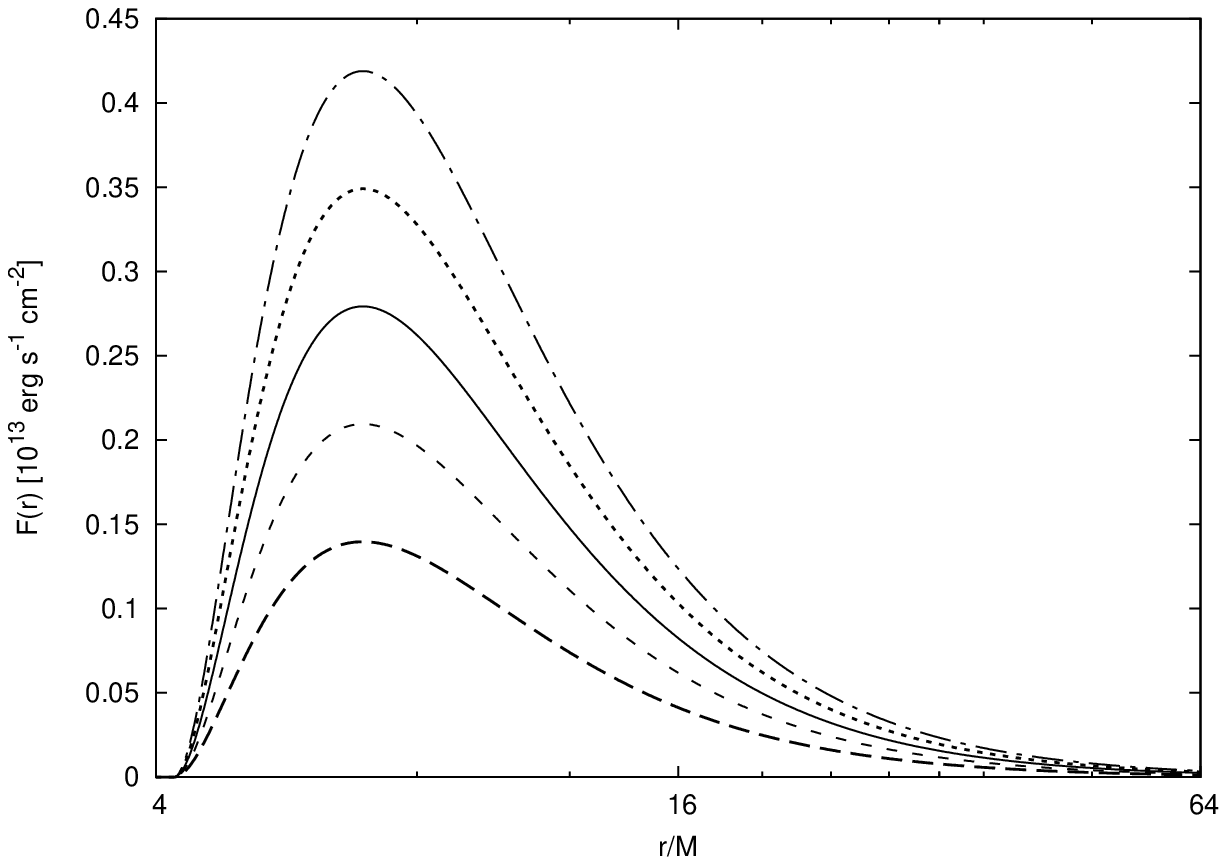}
\caption{Figure on the left: The time-averaged flux radiated by an accretion disk around a BMD
black hole for different values of $s$: $%
s=5$ (long dashed line), $s=10$ (short dashed line), $s=15$ (dotted line)
and $s=20$, respectively. The flux for a Schwarzschild black hole is plotted with a solid line. The mass accretion rate is $2\times 10^{-6}
M_{\odot}$ yr$^{-1}$. Figure on the right: The time-averaged flux radiated by the accretion disk around a BMD
black hole with $s=20$ for different values of the mass accretion rate $\dot{M}_0$: $1.0\times10^{-6} M_{\odot}$yr$^{-1}$ (long dashed line), $1.5\times10^{-6} M_{\odot}$yr$^{-1}$ (short dashed line), $2.0\times10^{-6} M_{\odot}$yr$^{-1}$ (solid line), $2.5\times 10^{-6} M_{\odot}$yr$^{-1}$ (dotted line), and $3.0\times 10^{-6} M_{\odot}$yr$^{-1}$ (dot-dashed line), respectively. The total mass of the black hole is $2.5\times 10^{6} M_{\odot}$.}
\label{BMD_F}
\end{figure}

With increasing $s$ the potential barrier increases as well, but we obtain
smaller and smaller radii for the marginally stable orbits. Although the
higher values of $s$ increase the potential over the region of the stable
Keplerian orbits, and the energy of the orbiting particles, as compared to the
case of the Schwarzschild potential, the value of $\sqrt{-g}$ used to
calculate the flux integral increases more rapidly. Therefore Eq.~(\ref{F})
gives smaller fluxes for higher values of $s$. This effect is shown in the left plot of Fig.~%
\ref{BMD_F}, where we present the plots of the photon flux emitted by the
accretion disk in the BMD brane black hole geometry.

\begin{figure}[tbp]
\includegraphics[width=8.15cm]{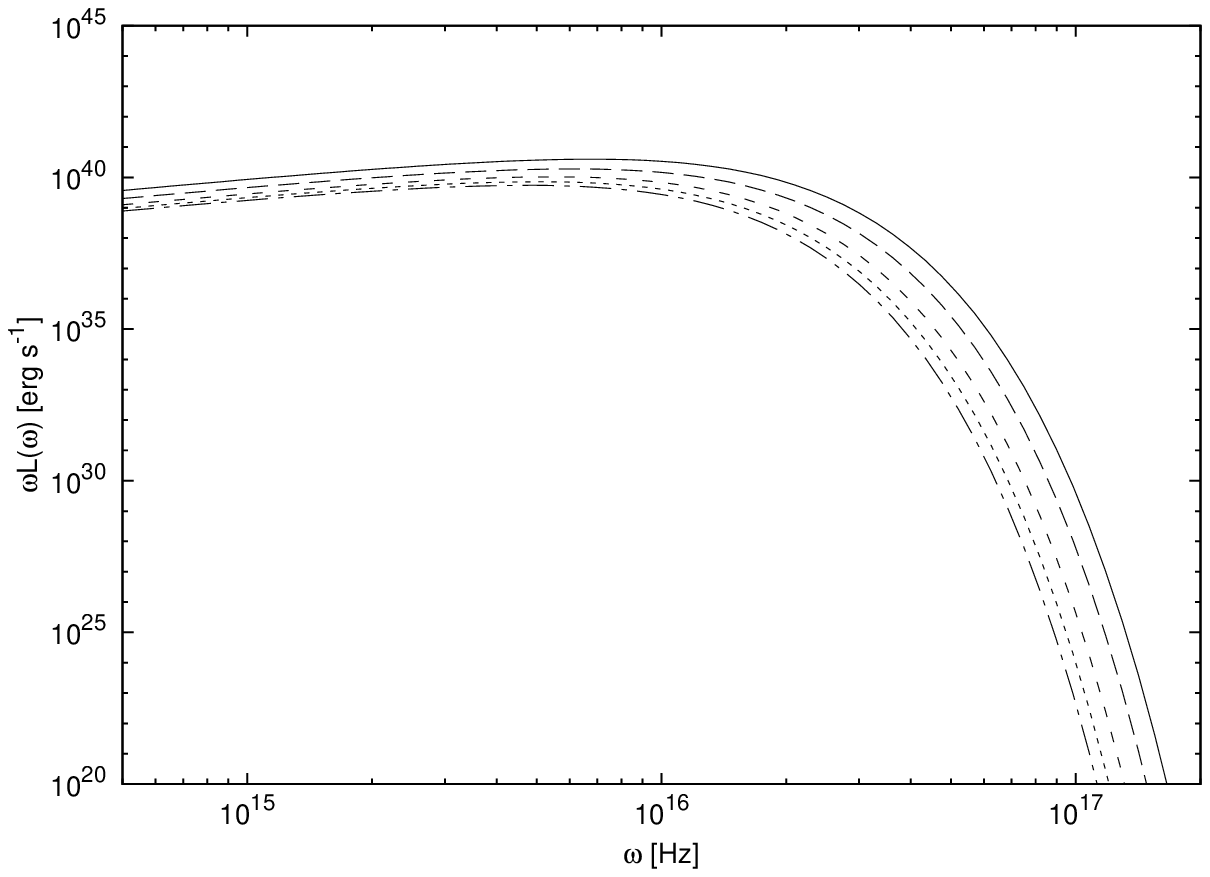}
\includegraphics[width=8.15cm]{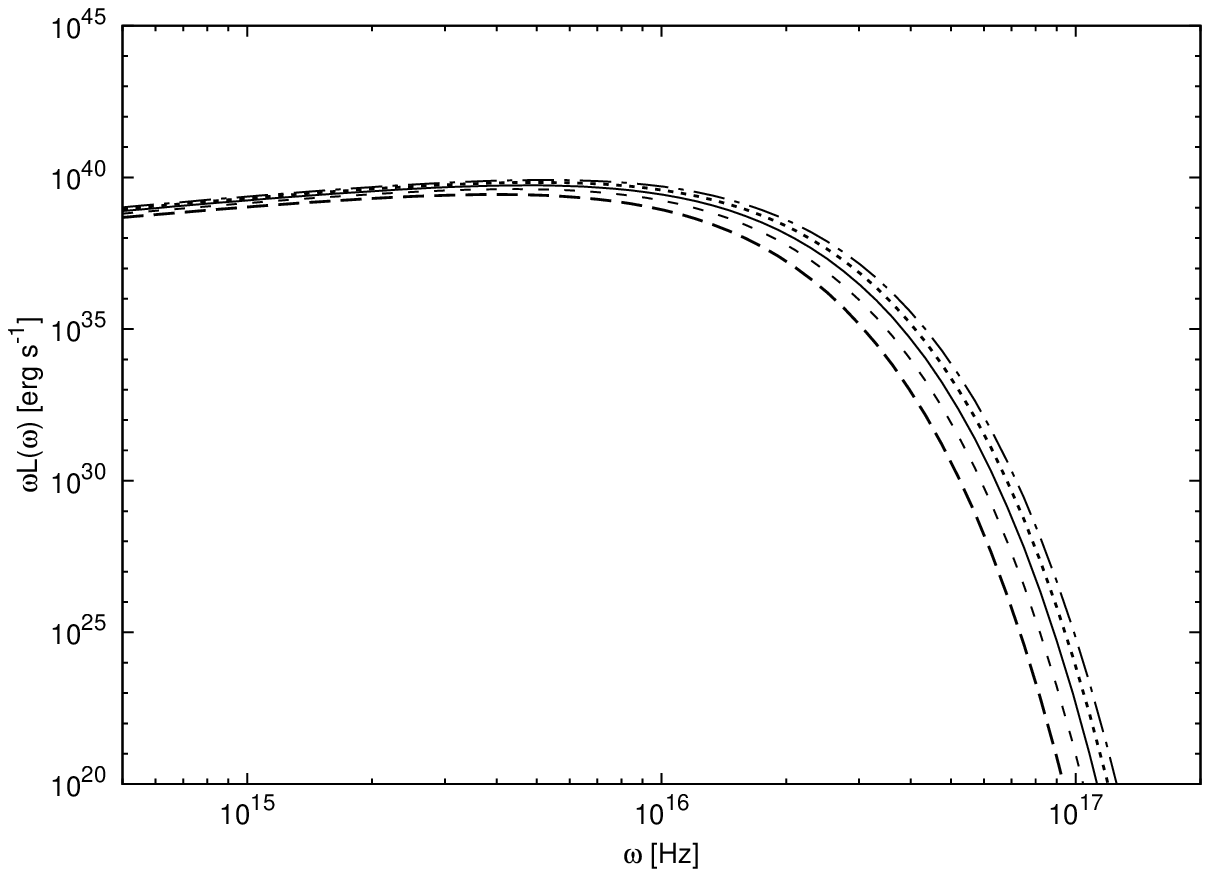}
\caption{Figure on the left: The emission spectrum of the thin accretion disk around a BMD black
hole for different values of the parameter $s$: $s=5$ (long dashed line), $s=10$ (short dashed line), $s=15$ (dotted
line), and $s=20$. The solid line represents the spectrum for the
Schwarzschild black hole. The mass accretion rate is $2\times 10^{-6}
M_{\odot}$ yr$^{-1}$. Figure on the right: The emission spectrum of the accretion disk around a BMD
black hole with parameter $s=20$ for different values of the mass accretion rate $\dot{M}_0$: $1.0\times 10^{-6} M_{\odot}$yr$^{-1}$ (long dashed line), $1.5\times10^{-6} M_{\odot}$yr$^{-1}$ (short dashed line), $=2.0\times10^{-6} M_{\odot}$yr$^{-1}$ (solid line),
$2.5\times 10^{-6} M_{\odot}$yr$^{-1}$ (dotted line), and $3.0\times 10^{-6} M_{\odot}$yr$^{-1}$ (dot-dashed line), respectively. The total mass of the black hole is $2.5\times 10^{6} M_{\odot}$.}
\label{BMD_L}
\end{figure}

The relative shift of $r_{ms}$ to lower orbits for increasing $s$ can
also be well studied in the plot. The maximum of the flux value has the
same shift: by increasing $s$ the maximum of $F(r)$ is obtained at lower and lower radii. This behavior of the radiation flux results in the softening of
the emission spectrum of the disk. As seen in the left plot of Fig.~\ref{BMD_L}, the cut-off
of the disk spectrum shifts to lower frequencies for increasing values of $s$%
, as compared to the ones obtained in the case of the Schwarzschild geometry. The right plots in Figs.~\ref{BMD_F} and \ref{BMD_L} show the effect of the change of the accretion rate for a BMD brane world black hole for a fixed mass an $s$.

The marginally stable orbit and the conversion efficiency $\epsilon$ of the
accreted mass into radiation measured at infinity for the BMD black holes
are presented in Table~ \ref{BMD_eps}. Both $r_{ms}$ and $\epsilon$ have
values less than those derived in the Schwarzschild geometry, and they
exhibit a decreasing tendency as we increase the values of the
parameter $s$.

\begin{table}[tbp]
\begin{center}
\begin{tabular}{|c|c|c|}
\hline
$s$ & $r_{ms}$ [$M$] & $\epsilon$ \\ \hline
5 & 4.8005 & 0.0303 \\ \hline
10 & 4.4007 & 0.0170 \\ \hline
15 & 4.2708 & 0.0118 \\ \hline
20 & 4.2046 & 0.0090 \\ \hline
\end{tabular}%
\end{center}
\caption{The marginally stable orbit and the efficiency for BMD black holes.}
\label{BMD_eps}
\end{table}

\subsection{The AG rotating brane world black hole}

The generalization of the Kerr solution of standard general relativity for
rotating brane world black holes was obtained by Aliev and  Gumrukcuoglu
\cite{AlGu05}, and is given by
\begin{eqnarray}
ds^{2}&=&-\left( 1-\frac{2Mr-\beta }{\Sigma }\right) dt^{2}-\frac{2a\left(
2Mr-\beta \right) }{\Sigma }\sin ^{2}\theta dtd\varphi +\frac{\Sigma }{%
\Delta }dr^{2}+\Sigma d\theta ^{2}+\nonumber\\
&&\left( r^{2}+a^{2}+\frac{2Mr-\beta }{%
\Sigma }a^{2}\sin ^{2}\theta \right) \sin ^{2}\theta d\varphi ^{2}.
\end{eqnarray}

We call this solution the AG rotating brane world black hole. The event
horizon of the black hole is determined by the solution of the equation $%
\Delta =0$, with the largest root given by $r_{+}=M+\sqrt{M^{2}-a^{2}-\beta }
$. The event horizon does exist if the condition $M^{2}\geq a^{2}+\beta $ is
fulfilled.  For a negative tidal charge, as $a\rightarrow M$, $%
r_{+}\rightarrow M+\sqrt{-\beta }>M$, a condition that is not allowed in
standard general relativity. On the other hand for a negative tidal charge
the extreme horizon $r_{+}=M$ corresponds to a black hole with rotation
parameter $a$ greater than its mass $M$.

\begin{figure}[tbp]
\includegraphics[width=8.15cm]{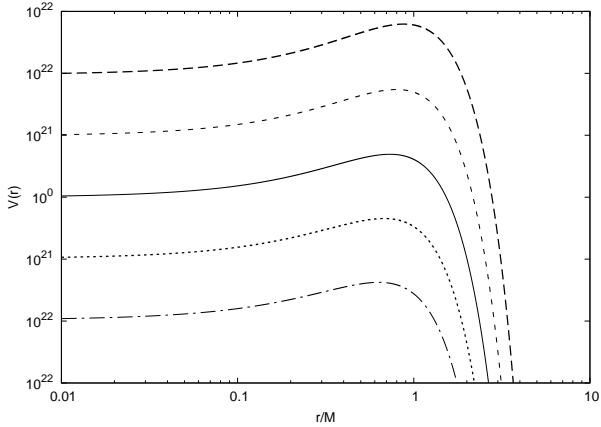}
\caption{The effective potential of a rotating AG brane world black hole of a total mass
$2.5\times 10^{6} M_{\odot}$ and with spin $a=0.9982$ for the specific energy $\widetilde{E}=0.8$ and the
specific angular momentum $\widetilde{L}=4M $. The solid line is the effective
potential for  a rotating Kerr black hole ($\beta=0$) with the same total mass. The different
values of $\beta$ are $\beta=-2M^2$ (long dashed line), $\beta=-M^{2}$ (short dashed line), $\beta=M^{2}$ (dotted
line) and $\beta=2M^{2}$ (dot-dashed line), respectively.}
\label{AG_V}
\end{figure}

Near to the equatorial plane, $|\theta-\pi/2|\ll1$, by introducing
the coordinate $z=r\cos\theta\approx r(\theta-\pi/2)$, the approximate form
for the geometry of the AG rotating brane black hole can be written as
\begin{equation}
ds^{2}=-\mathscr D\mathscr A^{-1}dt^{2}+r^{2}\mathscr A(d\phi-\omega dt)^{2}+\mathscr D^{-1}dr^{2}+dz^{2}\label{ds2AG}\end{equation}
with the metric functions \begin{eqnarray}
\mathscr A & = & 1+a_{*}^{2}/r_{*}^{2}+2a_{*}^{2}/r_{*}^{3}-a_{*}^{2}\beta_{*}/r_{*}
^{4}\;,\\
\mathscr D & = & 1-2/r_{*}+a_{*}^{2}/r_{*}^{2}-\beta_{*}/r_{*}^{2}\;\\
\omega& = & 2Mar^{-3}\mathscr{A}^{-1},
\end{eqnarray}
where we denoted $a_{*}=a/M$, $r_{*}=r/M$, and $\beta_{*}=\beta/M^{2}$, respectively. The effective potential per unit mass for the radial motion is given by
\begin{equation}
V(r)=\left[\left(r^2+a^2\right)r^2+\left(2Mr-\beta \right)a^2\right]\widetilde{E}^2-\left(r^2-2Mr+\beta \right)\widetilde{L}^2-2a\left(2Mr-\beta \right)\widetilde{E}\widetilde{L}-r^2\Delta.
\end{equation}
The variation of the potential $V(r)$ is represented, as a function of $r/M$, and for different values of the tidal charge parameter $\beta $, in Fig.~\ref{AG_V}.  By increasing $\beta $ from zero to $2M^2$ we also increase the potential barrier
as compared with the standard general relativistic Kerr black hole case, whereas negative tidal charges lowers the potential barrier. The changes in the value of $\beta $ also modify the positions of the
marginally stable orbits.

\begin{figure}[tbp]
\includegraphics[width=8.15cm]{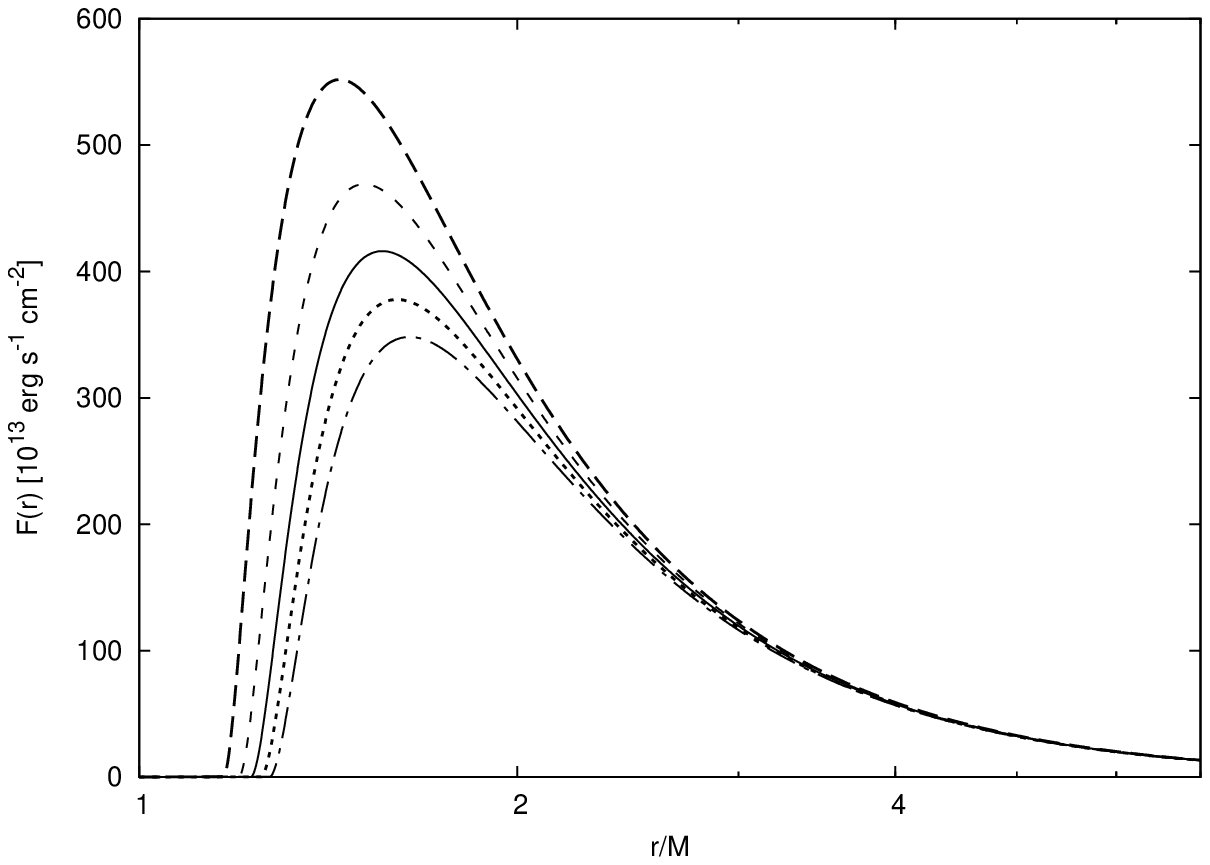}
\includegraphics[width=8.15cm]{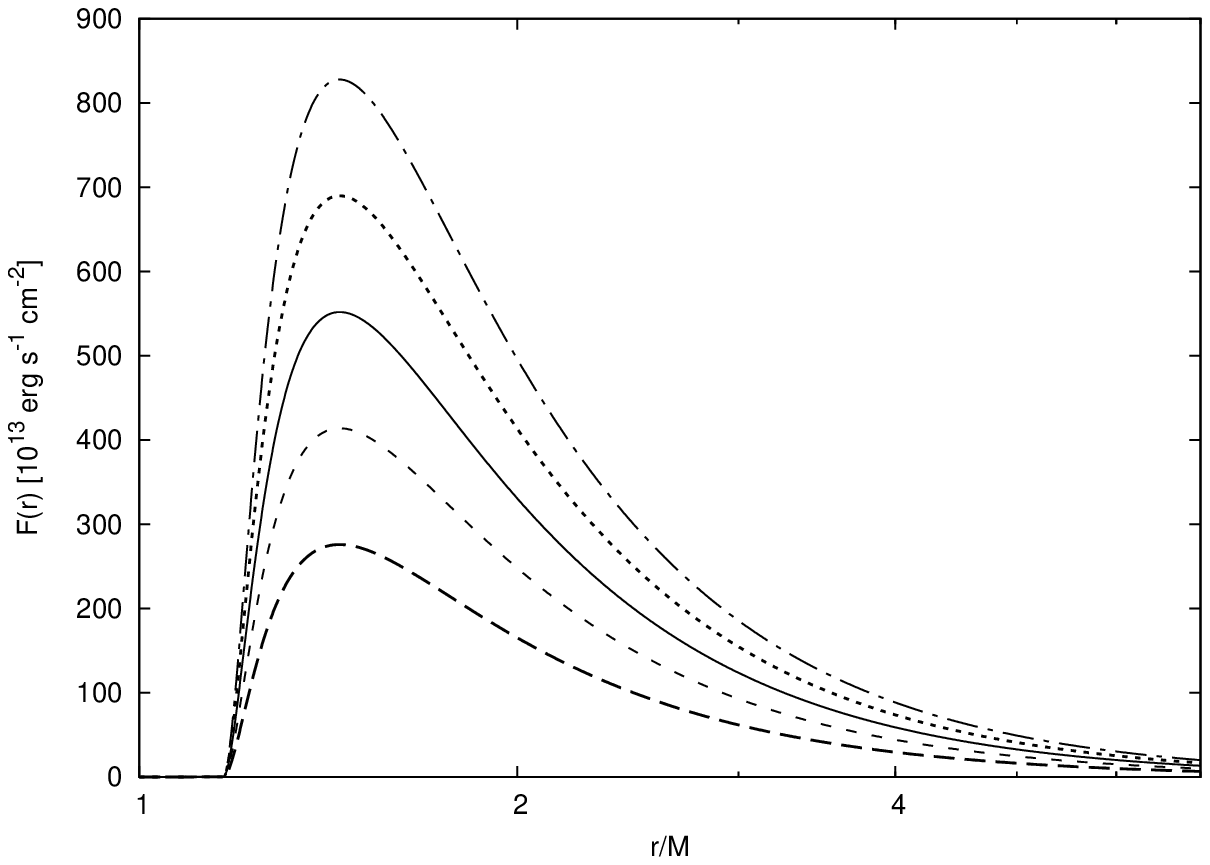}
\caption{Figure on the left: The time-averaged flux radiated by the accretion disk around a rotating AG brane world
black hole with spin $a=0.8892$ for different
values of the tidal charge parameter $\beta$: $\beta=-2\times10^{-3}M^2$ (long dashed line), $\beta=-10^{-3}M^{2}$ (short dashed line), $%
\beta=10^{-3}M^{2}$ (dotted line), and $\beta=2\times10^{-3}M^{2}$ (dot-dashed line), respectively. The flux for a Kerr black
hole with the same total mass and spin is plotted with a solid line. The
mass accretion rate is $2\times 10^{-6} M_{\odot}$yr$^{-1}$. Figure on the right: The time-averaged flux radiated by the accretion disk around a rotating AG brane world black hole with spin $a=0.8892$ and tidal charge $\beta=-2\times 10^{-3}M^2$ for different values of the mass accretion rate $\dot{M}_0$: $1.0\times10^{-6} M_{\odot}$yr$^{-1}$ (long dashed line), $1.5\times10^{-6} M_{\odot}$yr$^{-1}$ (short dashed line), $=2.0\times10^{-6} M_{\odot}$yr$^{-1}$ (solid line), $2.5\times 10^{-6} M_{\odot}$yr$^{-1}$ (dotted line), and $3.0\times 10^{-6} M_{\odot}$yr$^{-1}$ (dot-dashed line), respectively. The total mass of the black hole is $2.5\times10^6 M_{\odot}$. }
\label{AG_F}

\end{figure}

\begin{figure}[tbp]
\includegraphics[width=8.15cm]{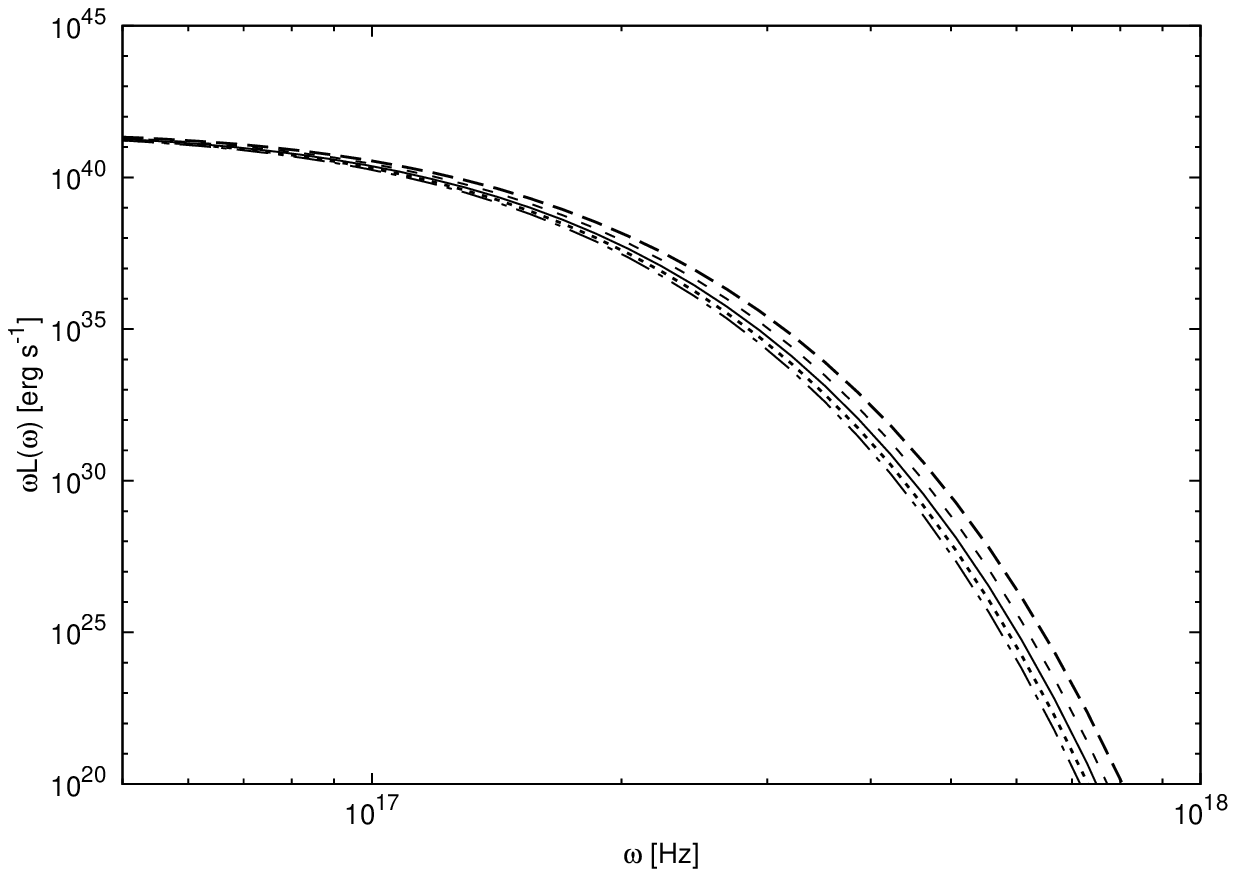}
\includegraphics[width=8.15cm]{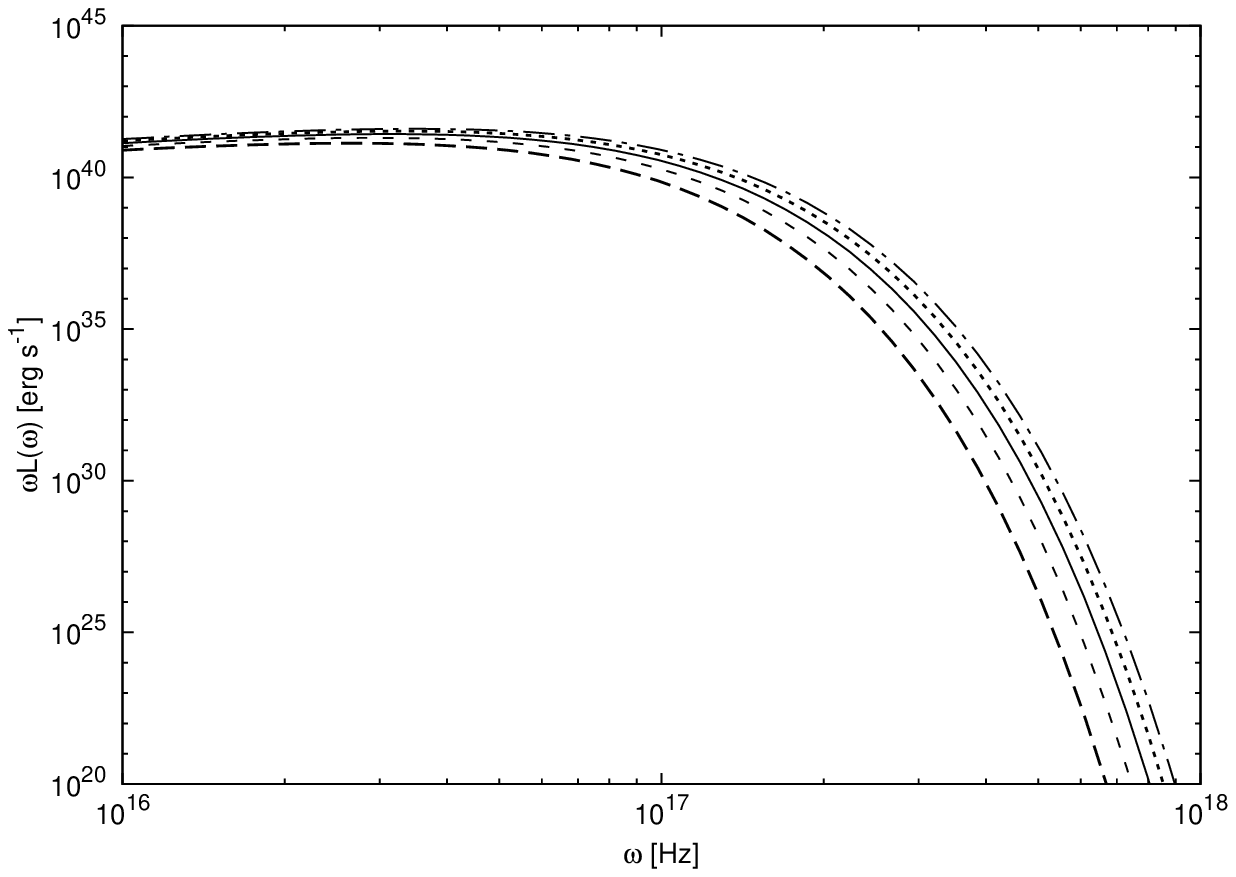}
\caption{Figure on the left: The emission spectrum of the accretion disk around a rotating AG brane world
black hole with spin $a=0.8892$ for different
values of the tidal charge parameter: $\beta=-2\times10^{-3}M^2$ (long dashed line), $\beta=-10^{-3}M^{2}$ (short dashed line), $%
\beta=10^{-3}M^{2}$ (dotted line), and $\beta=2\times10^{-3}M^{2}$ (dot-dashed line), respectively. The flux for a Kerr black
hole with the same total mass and spin is plotted with a solid line. The
mass accretion rate is $2\times 10^{-6} M_{\odot}$yr$^{-1}$. Figure on the right: The emission spectrum of the accretion disk around a rotating AG brane world black hole with spin $a=0.8892$ and tidal charge $\beta=-2\times10^{-3}M^2$ for different values of the mass accretion rate $\dot{M}_0$: $1.0\times10^{-6} M_{\odot}$yr$^{-1}$ (long dashed line), $1.5\times10^{-6} M_{\odot}$yr$^{-1}$ (short dashed line), $=2.0\times10^{-6} M_{\odot}$yr$^{-1}$ (solid line), $2.5\times 10^{-6} M_{\odot}$yr$^{-1}$ (dotted line), and $3.0\times 10^{-6} M_{\odot}$yr$^{-1}$ (dot-dashed line), respectively. The total mass of the black hole is $2.5\times10^6 M_{\odot}$.}
\label{AG_L}
\end{figure}

In Fig.~\ref{AG_F} we present the flux profiles of the accretion disk in the modified Kerr geometry (\ref{ds2AG}) as a function of $r/M$ and for different values of the tidal charge $\beta $ (left plot) and of the accretion rate $\dot{M}_0$ (right plot). The variation of the numerical value of the tidal charge determines  similar modifications for the flux values as in the case of the accretion disk around the DMPR black holes, which can be considered as the static limit ($a=0$) of the rotating AG black hole. The left hand plots in Figs.~\ref{AG_F} and \ref{DMPR_F}, respectively, show the same variation of the flux profiles as a function of the tidal charge. We note that since $Q$ corresponds to  $-\beta$, an increase in the numerical values of $\beta$ from negative values to positive ones decreases the magnitude of the flux, and increases the radius of the marginally stable orbits. The right hand plots in Figs.~\ref{DMPR_F}  and \ref{AG_F}, respectively,  also exhibit the same tendency: for higher mass accretion rates the flux will be amplified as well. In Figs.~\ref{DMPR_F} and \ref{AG_F} the cut-off values of the spectra decrease with the increasing values of $\beta$ and $-Q$, respectively. The same analogy is valid for disk spectra for the static and the rotating cases.  The emission spectra in the case of the rotating AG brane world black holes are represented, for different values of the spin parameter and of the accretion rate, in Figs.~\ref{AG_L}.


By comparing Tables~\ref{DMPR_eps} and \ref{AG_eps} we can see the same effect of the variation of the tidal charge
on the efficiency $\epsilon$ for the static DMPR and for the rotating AG brane black hole, respectively. As $\beta $ and $-Q$ increase from negative values to positive ones, the marginally stable orbits shift to higher radii, and the efficiency of the conversion of the accreting mass to radiant energy decreases, from values higher than 0.3241 (the efficiency for the standard Kerr black hole with $a_*=0.9982$) to lower ones.

\begin{table}[tbp]
\begin{center}
\begin{tabular}{|c|c|c|}
\hline
$\beta$ [$10^{-3}M^2$] & $r_{ms}$ [$M$] & $\epsilon$ \\ \hline
-2 & 1.1677 & 0.3449 \\ \hline
-1 & 1.2019 & 0.3329 \\ \hline
0 & 1.2277 & 0.3241 \\ \hline
1 & 1.2511 & 0.3169 \\ \hline
2 & 1.2716 & 0.3109 \\ \hline
\end{tabular}%
\end{center}
\caption{The marginally stable orbit and the efficiency for different rotating AG black hole geometries for $a_*=0.9982$. The case $\beta=0$ corresponds to the standard general relativistic Kerr black hole.}
\label{AG_eps}
\end{table}

\section{Discussions and final remarks}

In the present paper we have considered the basic physical properties of
matter forming a thin accretion disk in the space-time metric of the brane
world black holes. The physical parameters of the disk-effective potential,
flux and emission spectrum profiles have been explicitly obtained for
several classes of black holes, and for several values of the parameters
characterizing the vacuum solution of the generalized field equations in the
brane world models. All the astrophysical quantities related to the
observable properties of the accretion disk can be obtained from the black
hole metric.

There are many effective 4D solutions of the vacuum field equations on the
brane, with arbitrary parameters which depend on properties of the bulk, or
are simply put in by using general physical considerations. At the present
moment it is theoretically not known whether these parameters should be
universal over all brane world black holes, or whether each separate black
hole may have different values of them. Conversely, there is not a single
complete solution, in the sense that the metric in the bulk is uniquely known. This
situation is unsatisfactory from a purely theoretical point of view, and a
solution of this problem seems to be very difficult to be found. Therefore
it may be useful to solve the problem of the existence and nature of
the brane world black holes by investigating more closely the existing
observational evidence of the black hole properties, and try to discriminate between different black hole models by using the data provided by the observational study of the astrophysical processes around black holes.

Testing strong field gravity and the detections of the possible deviations from standard general relativity, signaling the presence of new physics, remains one of the most important objectives of observational astronomy. Due to their compact nature, black holes provide an ideal environment to do this. Presently, the best constraints on the brane world black hole parameters can be obtained from the classical tests of general relativity (perihelion precession, deflection of light, and the radar echo delay, respectively). The existing observational solar system data on the perihelion shift of Mercury, on the light bending around the Sun (obtained using long-baseline radio interferometry), and ranging to Mars using the Viking lander, as applied to the DMPR black hole, can constrain the numerical values of both the bulk tidal parameter $Q$ and of the brane tension \cite{test}. The stronger limit is obtained from the perihelion precession, $\left\vert Q\right\vert \leq 6\times 10^{7}-5\times 10^{8}$ cm$^{2}$. An improvement of one order of magnitude in the observational data on Mercury's perihelion shift could provide a very precise estimate of the bulk tidal parameter, as well as of the brane tension $\lambda $.

Observations in the near-infrared (NIR) or X-ray bands have provided important information about the spin of the black holes, or the absence of a surface in stellar type black hole candidates. In the case of the source Sgr A$^*$, where the putative thermal emission due to the small accretion rate peaks in the near infrared, the results are particularly robust. However, up to now, these results have confirmed the predictions of the general relativity mainly in a qualitative way, and the observational precision achieved cannot distinguish between the different proposed theories of gravitation. However, important technological developments may allow to image black holes directly \cite{BrNa06a}. A background illuminated black hole will appear in a silhouette with radius $\sqrt{27}GM/c^2$, with an angular size of roughly twice that of the horizon, and may be directly observed. With an expected resolution of $20$ $\mu $as, submillimeter very-long baseline interferometry (VLBI) would be able to image the silhouette cast upon the accretion flow of Sgr A$^*$, with an angular size of $\sim 50$ $\mu $as, or M87, with an angular size of $\sim 25$ $\mu $as. For a black hole embedded in an accretion flow, the silhouette will generally be asymmetric regardless of the spin of the black hole. Even in an optically thin accretion flow asymmetry will result from special relativistic effects (aberration and Doppler shifting). In principle, detailed measurements of the size and shape of the silhouette could yield information about the mass and spin of the central black hole, and provide invaluable information on the nature of the accretion flows in low luminosity galactic nuclei.

Due to the differences in the space-time structure, the brane world black
holes present some important differences with respect to their disc accretion
properties, as compared to the standard general relativistic Schwarzschild and Kerr
cases. Therefore, the study of the accretion processes by compact objects is
a powerful indicator of their physical nature. Since the conversion
efficiency, as well as the flux and the spectrum of the black body radiation
in the case of the brane world black holes is different as compared to the
standard general relativistic case, the astrophysical determination of these
physical quantities could discriminate, at least in principle, between the
different gravity theories, and give some constrains on the existence of the
extra dimensions.

\section*{Acknowledgments}

The work of T. H. is supported by an RGC grant of the government of the Hong
Kong SAR.

\end{document}